\documentclass[prx,twocolumn,floatfix,superscriptaddress,longbibliography]{revtex4-1}
\usepackage{tikz}
\usepackage{mathtools}
\usetikzlibrary{shapes,arrows,positioning}
\usepackage{standalone}
\usepackage{amssymb,amsmath,amstext}
\usepackage{graphicx}
\usepackage{epstopdf}
\usepackage{color}
\usepackage{appendix}
\usepackage[T1]{fontenc}
\usepackage{bbold}
\usepackage{bbm}
\usepackage{float}
\usepackage{latexsym}
\usepackage{xr-hyper}
\usepackage[colorlinks=true,citecolor=blue,linkcolor=magenta]{hyperref}
\usepackage[latin1]{inputenc}
\usepackage[linesnumbered,lined,boxed,commentsnumbered]{algorithm2e}
\usepackage{cancel}

\usepackage{tikz}
\usetikzlibrary{shapes,arrows}
\renewcommand{\vec}[1]{\boldsymbol{#1}}  

\long\def\ca#1\cb{} 

\newcommand{\ket}[1]{|#1\rangle}               
\newcommand{\bra}[1]{\langle #1|}              
\newcommand{\dya}[1]{\ket{#1}\!\bra{#1}}

\newcommand{\AC}{\mathcal{A}}

\newcommand{\EC}{\mathcal{E}}
\newcommand{\FC}{\mathcal{F}}

\newcommand{\IC}{\mathcal{I}}

\newcommand{\UC}{\mathcal{U}}

\newcommand{\Tr}{{\rm Tr}}

\newcommand{\mte}[2]{\langle#1|#2|#1\rangle }

\newcommand{\UCC}{\text{UC}}
\newcommand{\SP}{\text{SP}}
\newcommand{\OBE}{\text{OE}}
\newcommand{\alv}{\vec{\alpha}}
\newcommand{\thv}{\vec{\theta}}
\newcommand{\kv}{\vec{k}}
\newcommand{\Fbar}{\overline{F}}

\renewcommand{\vec}[1]{\boldsymbol{#1}}  
\newcommand{\ot}{\otimes}
\newcommand{\ad}{^\dagger}

\newcommand*{\id}{\openone}

\newcommand{\Par}{\text{Par}}

\newcommand{\etal}{\emph{et al.}~}
\newcommand{\ie}{\emph{i.e.,}~}
\newcommand{\eg}{\emph{e.g.,}~}
\newcommand{\nn}{\nonumber}

\begin{document}

\title{Machine learning of noise-resilient quantum circuits}

\author{Lukasz Cincio} 
\affiliation{Theoretical Division, MS 213, Los Alamos National Laboratory, Los Alamos, NM 87545, USA.}

\author{Kenneth Rudinger}
\affiliation{Quantum Computer Science, Sandia National Laboratories, Albuquerque, NM 87185, USA}

\author{Mohan Sarovar}
\email{mnsarov@sandia.gov}
\affiliation{Extreme-scale Data Science and Analytics, Sandia National Laboratories, Livermore, CA 94550, USA}

\author{Patrick J. Coles} 
\email{pcoles@lanl.gov}
\affiliation{Theoretical Division, MS 213, Los Alamos National Laboratory, Los Alamos, NM 87545, USA.}

\begin{abstract}
Noise mitigation and reduction will be crucial for obtaining useful answers from near-term quantum computers. In this work, we present a general framework based on machine learning for reducing the impact of quantum hardware noise on quantum circuits. Our method, called noise-aware circuit learning (NACL), applies to circuits designed to compute a unitary transformation, prepare a set of quantum states, or estimate an observable of a many-qubit state. Given a task and a device model that captures information about the noise and connectivity of qubits in a device, NACL outputs an optimized circuit to accomplish this task in the presence of noise. It does so by minimizing a task-specific cost function over circuit depths and circuit structures. To demonstrate NACL, we construct circuits resilient to a fine-grained noise model derived from gate set tomography on a superconducting-circuit quantum device, for applications including quantum state overlap, quantum Fourier transform, and $W$-state preparation.
\end{abstract}
\maketitle

\section{Introduction}
Recent years have seen a surge in quantum computer hardware development, and we now have several quantum computing platforms with tens of qubits that can be controlled and coupled with fidelities that enable execution of quantum circuits of limited depth. This has led to intense interest in formulating quantum algorithms that can be reliably executed on such devices. The challenge however is that naive compilations of nearly all non-trivial quantum algorithms require circuit depths that are currently out of reach for near-term hardware. Motivated by this challenge, in this work we study how machine learning (ML) can be applied to formulate noise-aware quantum circuits that can be executed on near-term quantum hardware to produce reliable results. 

\begin{figure}
\includegraphics[width=0.75\columnwidth]{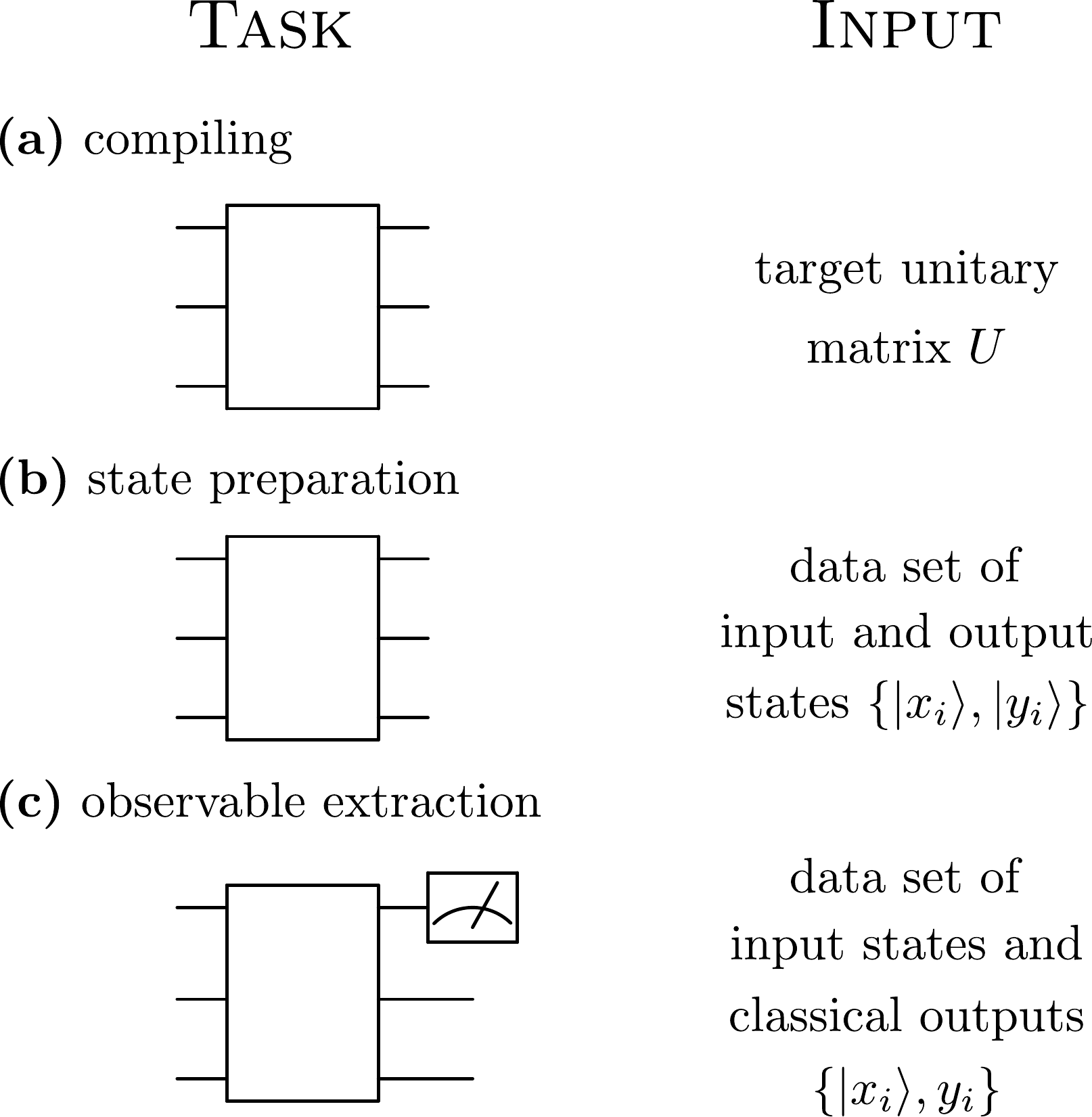}
\caption{Applications of NACL. (a) In compiling, the goal is to approximate an input unitary matrix $U$ by a noise-resilient circuit that is compatible with the device constraints. (b) In state preparation, one inputs a set of $N$ input and output states $\{ \ket{x_i}, \ket{y_i}\}$, where $N$ could be as small as one, and the output is a noise-resilient circuit that approximately prepares the $\ket{y_i}$ states from the $\ket{x_i}$ states. (c) In observable extraction, one inputs a set of input states and classical outputs that typically correspond to local observable expectation values, $\{ \ket{x_i}, y_i\}$, and the output is a noise-resilient circuit that approximately computes the outputs from any input state $\ket{\psi}$ that might or might not be in the input set.
}
\label{fig:Applications}
\end{figure}

Our method is called noise-aware circuit learning (NACL), and given suitable description of a computational task and a device model that captures the noise and constraints of a device, it outputs a native circuit that performs the task with greatest robustness to noise. NACL has several broad applications, as illustrated in Fig.~\ref{fig:Applications}. The task can be the compilation of a specified unitary transformation (Fig.~\ref{fig:Applications}(a)), the preparation of a target state from a specified input state (Fig.~\ref{fig:Applications}(b)), or the extraction of an observable from a many-qubit state (Fig.~\ref{fig:Applications}(c)). In each case, NACL returns a circuit that is the significantly more noise-resilient to the given noise model, however, as we detail below, the formulation of the machine learning problem is different in each application. Perhaps the most familiar version of NACL is that depicted in Fig.~\ref{fig:Applications}(a), where a specified unitary matrix is to be implemented by a circuit composed of native gates, which is usually called compilation. In this context, NACL results in noise-aware circuit compilations.

Previous work on circuit optimization for noise mitigation has largely considered the task of compilation, under restricted models of errors or imperfections. In fact, most work focuses on reducing overall circuit error by reducing the number of two-qubit gates (which tend to be more noisy than single-qubit gates), avoiding faulty qubits, reducing the number of SWAP gates required in architectures with restricted connectivity, or reducing the amount of qubit idle time and/or overall circuit depth \cite{tucci_qubiter_2004,javadiabhari_scaffcc:_2014,maslov_basic_2017,venturelli_compiling_2017,murali_formal_2019, cincio2018learning, tannu_not_2019, sivarajah_tketrangle_2020}. These strategies incorporate very little information about errors present in a particular hardware platform. More recent work on error-aware compilation by Murali \etal 
\cite{murali_noise-adaptive_2019} goes beyond this and includes basic calibration information (\eg qubit $T_2$ times, CNOT gate error rates) to compile circuits using more reliable qubits and gates.

In this work we extend this direction even further and demonstrate that one can use fine-grained error model information to increase the reliability of the outputs of quantum circuits. Incorporating detailed noise models into one's circuit optimization, as we do here, is particularly compelling at present with the advent of advanced characterization techniques like gate-set tomography \cite{blume-kohout_demonstration_2017,noauthor_pygsti._nodate}. These techniques produce fine-grained details -- \eg estimates of process matrices representing the action of imperfect quantum gates -- describing the actual evolution of qubits in near-term hardware. We will demonstrate that such experimentally derived noise models can be used to go beyond naive circuit compilations for several example quantum algorithms.

NACL has several additional strengths relative to existing approaches in the literature. Crucially, NACL takes a task-oriented approach to quantum circuit discovery, which implies that one does not need a starting point or example quantum circuit that already accomplishes the task. Note that traditional compilers do require such a quantum circuit to start from. Furthermore, because NACL does not start from a template circuit, the optimization is less susceptible to bias. In contrast, standard literature methods that tweak a given quantum circuit inherently bias their optimization towards solutions that look like that starting point. This means that NACL has the potential to discover more novel solutions that otherwise would not be obvious to the human mind. In addition, we will see that NACL naturally balances the trade-off between circuit depth, which leads to more expressivity, and circuit noise, which makes outputs less accurate.

Machine learning was previously applied to train parameterized quantum circuits~\cite{cincio2018learning, wan2017quantum}, albeit in a noise-free setting. In addition, variational quantum algorithms (VQAs)~\cite{peruzzo2014VQE, mcclean2016theory, farhi2014QAOA, Romero17, li2017efficient, mitarai2018quantum, Khatri2019quantumassisted, jones2018quantum, larose2018, arrasmith2019variational, cerezo2020variational, bravo-prieto2019,cirstoiu2019variational,cerezo2020variational2} can also be thought of as machine learning of quantum circuits. In the Discussion (Sec.~\ref{sec:disc}), we elaborate on the relationship between NACL and VQAs.

In what follows, we first present our theoretical framework (Sec. \ref{sec:Framework}). We then discuss a device model with experimentally determined noise parameters (Sec.~\ref{sec:nm}). Next, we present our implementations of NACL with this noisy device, for examples from the three different application classes shown in Fig.~\ref{fig:Applications} (Secs. \ref{sec:impl_obs} - \ref{sec:impl_comp}). Finally, we conclude with a discussion in Sec. \ref{sec:disc}.

\section{Machine Learning Framework}\label{sec:Framework}

\subsection{Overview}\label{sec:Overview}

\begin{figure}
\includegraphics[width=\columnwidth]{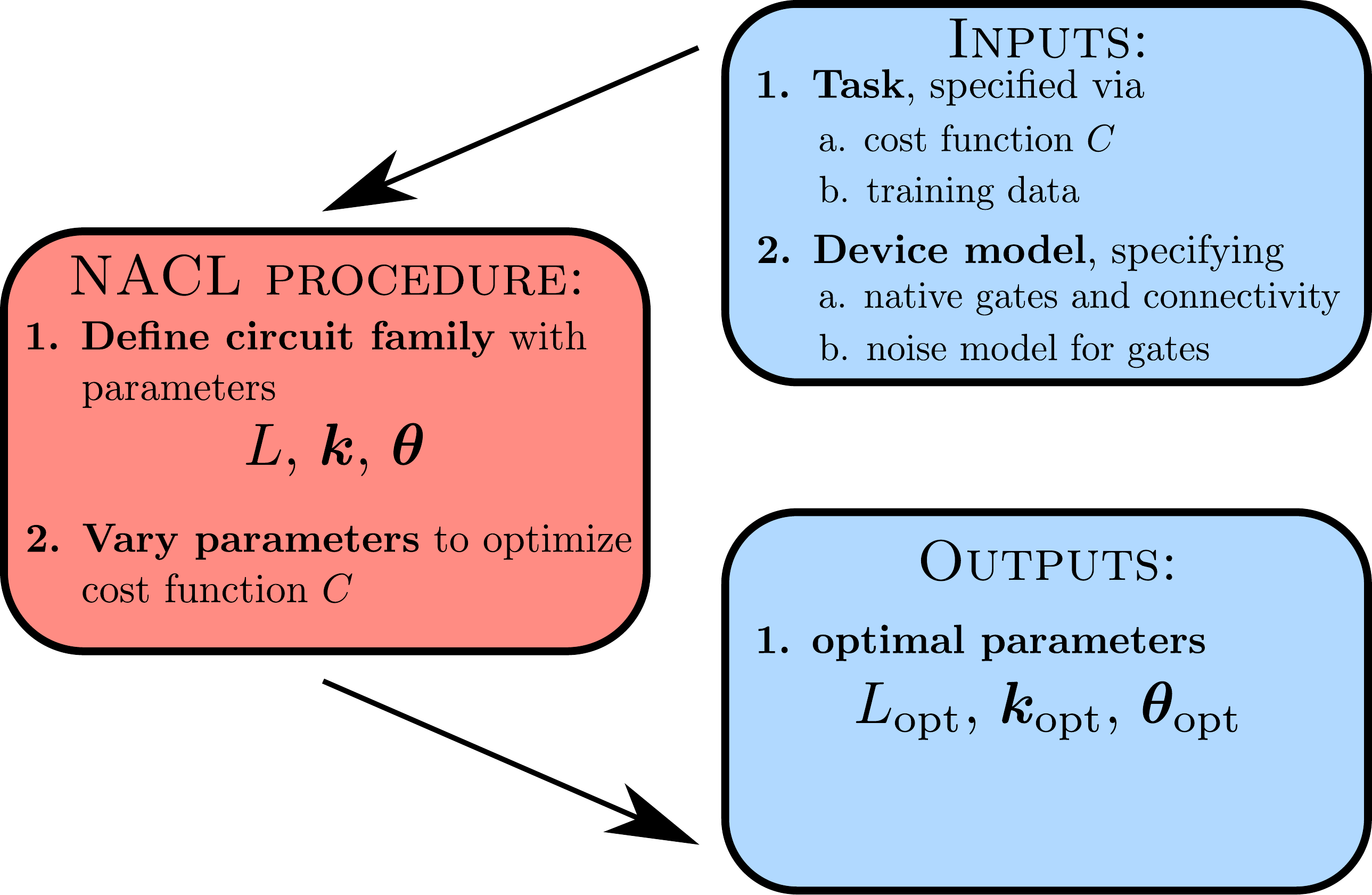}
\caption{Schematic diagram of NACL. Our approach takes a task and a device model as an input. The task is defined via examples in a training set and a cost function, $C$. That information is sufficient to find a noise-aware circuit that approximates a specified task. It is done via optimization over a set of parameters $(L,\vec{k},\vec{\theta})$ that describe a quantum circuit. The algorithm returns parameters $(L_\mathrm{opt},\vec{k}_\mathrm{opt},\vec{\theta}_\mathrm{opt})$, which represent an optimized quantum circuit that minimizes the cost function $C$. See text for details.  
}
\label{fig:Schematic}
\end{figure}

A schematic diagram of the steps of NACL is shown in Figure~\ref{fig:Schematic}. There are two inputs to NACL: (1) a task, and (2) a device model. The output of NACL is an optimized quantum circuit that accomplishes the inputted task in the presence of the inputted device model. NACL may not output a globally optimal solution (this depends on details of the cost function landscape and optimization method used), but even local optima are improvements over circuit compilations that are not noise-aware.

Note that circuit depth is not an input to NACL. This is because NACL optimizes over circuit depths, and aims to find the depth that achieves the most noise resilience. In addition, an ansatz for the circuit is not an input, because NACL attempts to optimize over many ans\"atze. Hence, the structure of the circuit, as well as its depth, are optimized by NACL. This feature of NACL is in the spirit of task-oriented programming, where the user only needs to specify the task, and not the details of the circuit. NACL adapts the circuit structure to optimize a cost function that depends on the type of task specified. As shown in Fig. \ref{fig:Applications}, there are three categories of tasks. 

In what follows we provide more details on how NACL works. Sections~\ref{sec:GateSequence} and \ref{sec:NoiseModels} discuss the device model and noise specification, and Section~\ref{sec:CostFunctions} defines the NACL cost function for each application. Finally, Section~\ref{sec:Optimization} summarizes the optimization methods used by NACL.

\subsection{Parameterized circuit}\label{sec:GateSequence}

For a given quantum hardware, we denote the native gate set or gate alphabet as $\AC = \{ A_j (\theta)\}$. Each gate $A_j$ is either a one- or two-qubit gate and may also have an internal continuous parameter $\theta$. As an example, the IBM Q 5-qubit computer ``Ourense'' has the native gate alphabet
\begin{align}
\label{eq:ouresne_gs}
\AC_{\text{Ourense}} = \{ & \text{CNOT}^{12}, \text{CNOT}^{23}, \text{CNOT}^{24}, \text{CNOT}^{45}, \notag\\
& Z^1(\theta) , X^1(\pi/2), Z^2(\theta) , X^2(\pi/2) , \notag \\
& Z^3(\theta) , X^3(\pi/2), Z^4(\theta) , X^4(\pi/2) , \notag \\
& Z^5(\theta) , X^5(\pi/2) \}\,,
\end{align}
where $\text{CNOT}^{jk}$ is a $\text{CNOT}$ between qubits $j$ and $k$, $Z^j(\theta)$ is a rotation of angle $\theta$ about the z-axis of qubit $j$, and $X^j(\pi/2)$ is a rotation of angle $\pi/2$ about the x-axis of qubit $j$ (also called a pulse gate). 

Such a gate set is supplemented by state preparation and measurement quantum operations. These are typically fixed in most quantum computing architectures (\eg prepare all qubits in the ground state and measure in the computational basis), and therefore there is no opportunity for optimizing over these. Therefore, we do not consider these as part of the learnable set.

We consider a generic gate sequence that defines a circuit
\begin{equation}
\label{Galpha}
G_{\alv} = G_{(L, \kv, \thv)} =   A_{k_L}(\theta_{L}) \cdot\cdot\cdot A_{k_2}(\theta_2) A_{k_1}(\theta_1)\,,
\end{equation}
where $L$ is the number of gates, $\vec{k} = (k_1,..., k_L)$ is the vector of indices describing which gates are utilized in the gate sequence, $\vec{\theta} = (\theta_1,..., \theta_L)$ is the vector of continuous parameters associated with these gates, and $\alv = (L, \kv, \thv)$ is the set of all these parameters. All parameters in $\alv = (L, \kv, \thv)$ are optimized over in NACL.

\subsection{Device model}\label{sec:NoiseModels}

An input to NACL is a device model, which captures the constraints of a device (\eg limited connectivity) and also represents the noise in the device. We assume the device constraints and connectivity are captured by the specification of a native gate alphabet for the device, \eg Eq. \eqref{eq:ouresne_gs}. Only gates that are available are listed in this specification. 

The salient characteristics of noise are captured by (i) process matrices for each element of the device's native gate alphabet, and (ii) for state preparation and measurement (SPAM) noise, by quantum-classical channels that represent noisy state preparation or measurement POVM elements. The assumption of a fixed process matrix for each gate in the alphabet restricts this treatment to Markovian noise. This can be relaxed by generalizing to time-dependent process matrices for each elementary gate, but we do not do this here for simplicity, and also because characterization tools capable of producing such non-Markovian representations of quantum computer operations are still in early stages of development \cite{proctor_detecting_2019}. Similarly, in this treatment we mostly ignore the effects of crosstalk, and assume that the process matrix describing a gate operates only on the qubits the ideal gate is defined on. Properly incorporating crosstalk into the noise models that NACL considers requires advances in characterization methods \cite{sarovar_detecting_2019} that we discuss later.

Given this paradigm for representing noisy quantum operations, each gate in the alphabet $\AC$ has an associated process matrix that accounts for the local noise occurring during that gate. Note that even the identity gate may have a non-trivial process matrix, for example due to relaxation during idling. 

Mathematically speaking, the noise model provides a map from a parameterized circuit $G_{\alv}$ to a parameterized quantum channel $\EC_{\alv}$:
\begin{equation}
\label{eqn3}
    G_{\alv}\xrightarrow[]{\text{Noise Model}} \EC_{\alv}\,.
\end{equation}
Here, $\EC_{\alv}$ is a completely positive trace preserving (CPTP) map that represents the action of $G_{\alv}$ in a noisy environment.

Specifically, when the noise model is given in the form of process matrices for gates, one can do the following. Let $\AC = \{ A_j (\theta)\}$ denote the gate alphabet associated with the noiseless gates. In the presence of noise, this gate alphabet becomes a set of quantum channels, $\bar{\AC} = \{ \bar{A}_j (\theta)\}$, where we note that $\bar{A}_j (\theta)$ now denotes a quantum channel. Now suppose that $G_{\alv}$ is given by $G_{\alv} =   A_{k_L}(\theta_{L}) \cdot\cdot\cdot A_{k_2}(\theta_2) A_{k_1}(\theta_1)$. Then the simplest way to incorporate the noise model would be to replace each $A_{k_i}$ with $\bar{A}_{k_i}$; \ie to transform $G_{\alv}$ into a sequence of quantum channels:
\begin{equation}
    \EC_{\alv} = \bar{A}_{k_L}(\theta_{L}) \circ \dots \circ \bar{A}_{k_2}(\theta_2) \circ \bar{A}_{k_1}(\theta_1)\,.
\end{equation}
However, it is important to note that this formula for $\EC_{\alv}$ only accounts for the non-trivial gates that were in the original circuit $G_{\alv}$. However, in practice, identity gates will occur with noise due to, \eg thermal relaxation. Therefore, care must be taken with respect to identity gates, and we discuss this next.

\subsubsection{Parallelization}

The object we are optimizing over, the circuit in Eq. \eqref{Galpha}, needs to be modified in the presence of imperfect idle operations. In this case, the sensible thing to do is to perform as many gates in parallel as possible, but the description of a circuit as a sequence of gates, as in Eq.~\eqref{Galpha}, is incomplete because it does not capture which gates can be performed in parallel. In other words, in the presence of imperfect idle operations we cannot simply think of $G_{\alv}$ as a linear sequence of gates; we have to map $G_{\alv}$ to a two-dimensional circuit diagram, in space and time.

Abstractly, we can re-write 
\begin{equation}
\label{Galphapar}
G_{\alv}^{\Par} = G_{\alv} = U_{\alv, M} \cdot\cdot\cdot U_{\alv, 2} U_{\alv, 1} \,.
\end{equation}
Here, each $U_{\alv, j}$ represents a layer of gates that can be parallelized. Specifically, we take the circuit proposed in $G_{\alv}$ and compress it using simple circuit rules to minimize idling of qubits. For example, an $X(\pi/2)$ rotation that occurs on the target qubit after a CNOT can be moved to before the CNOT because their actions on the target qubit commute. In this manner, each gate in $G_{\alv}$ is moved to as early a time as possible without changing the unitary being implemented by $G_{\alv}$. This naturally defines the circuit layers and subsequently $G_{\alv}^{\Par}$. Even though the reordering does not change the overall unitary, whenever we write $G_{\alv}$ in the form in Eq.~\eqref{Galphapar} we denote it as $G_{\alv}^{\Par}$. An important aspect of the optimization in NACL is to numerically find the parallelized representation, $G_{\alv}^{\Par}$, that yields the minimum error in the cost functions detailed below. 

Once $G_{\alv}$ is rewritten in the form of $G_{\alv}^{\Par}$, we can then account for noise by replacing each gate in $G_{\alv}^{\Par}$ by the quantum channel that represents its noisy implementation. For example, if a circuit layer in $G_{\alv}^{\Par}$ on a 5-qubit processor happens to be
\begin{align}
    U_{\alv, j} = Z^1(\theta) \otimes \textrm{CNOT}^{23} \otimes I^4 \otimes X^5,
\end{align}
where the superscript indicates which qubit the gates are operating on (and $Z(\theta)$ is a rotation around the $Z$ axis, CNOT is a CNOT gate, $I$ is the identity, and $X$ is a $\pi/2$ rotation around the $X$ axis).
This layer would be replaced by
\begin{align}
    \bar{U}_{\alv, j} = \bar{Z}^1(\theta) \otimes \overline{\textrm{CNOT}}^{23} \otimes \bar{I}^4 \otimes \bar{X}^5,
\end{align}
where the quantities with bars above them are the quantum channels representing those gates. Then the overall noisy circuit corresponding to $G_{\alv}^{\Par}$ is written as
\begin{equation}
\label{eqn7}
\EC_{\alv}^{\Par}  = \bar{U}_{\alv, M} \circ \dots \circ \bar{U}_{\alv, 2} \circ \bar{U}_{\alv, 1} .
\end{equation}
It is important to note that NACL uses $\EC_{\alv}^{\Par}$ rather than $\EC_{\alv}$ as the overall noisy channel associated with $G_{\alv}$.

Note that this procedure of parallelizing and incorporating the noise model that we have outlined is valid because our noise models do not account for crosstalk effects. If crosstalk is significant, then this strategy of maximizing parallelization might not be optimal since performing many gates in parallel may lead to more noise. Moreover, in the presence of significant crosstalk, capturing processor noise using quantum channels for each of its gates is probably insufficient. Instead, one would need to characterize each possible layer (there are an exponential number of these) since the operation on a qubit due to application of a gate could depend on what is done to any other qubit in the computer at the same time. We discuss how to extend NACL in the presence of crosstalk in the Sec. \ref{sec:disc}.

\subsection{Cost functions}\label{sec:CostFunctions}
In this subsection, we construct the cost functions that are minimized by NACL in each of the application classes outlined in Fig. \ref{fig:Applications}. 

\subsubsection{Preliminaries}

We first define some relevant quantities. Let $F(\rho , \sigma)  = (\Tr \sqrt{\sqrt{\rho} \sigma \sqrt{\rho} })^2 $ be the fidelity between two states $\rho$ and $\sigma$. For a given pure input state $\ket{\psi}$, we can denote the fidelity of the output states under quantum channels $\EC$ and $\FC$ as
\begin{align}
    F(\EC, \FC, \ket{\psi}) := F(\EC(\dya{\psi}) , \FC(\dya{\psi}))\,.
\end{align}
We will be interested in the case where $\FC$ corresponds to a unitary process $\UC$, in which case we have
\begin{align}
    F(\EC, \UC, \ket{\psi}) = \Tr(\EC(\dya{\psi})  \UC(\dya{\psi}))\,.
\end{align}
Furthermore, we can define the average process fidelity as
\begin{align}
    \Fbar (\EC , \UC) &= \int d\psi F(\EC, \UC, \ket{\psi})\\
    &= \int d\psi \Tr(\EC(\dya{\psi})  \UC(\dya{\psi}))\,,
    \label{eq:Fbar_process}
\end{align}
with the integral taken over the Haar measure.

\subsubsection{Observable extraction}

The first class of applications involves estimating an observable given one, or a set of, input states. An example of this is computing the overlap of two quantum states (discussed in Section~\ref{sec:impl_obs}). In this application, the output of the circuit is a classical number (the observable expectation, which in practice is estimated by many executions of the circuit) denoted $f(x)$, and the input, denoted $\ket{x}$, is a quantum state (or classical data encoded in a quantum state). Hence, we want to construct a circuit that computes the function $\ket{x} \rightarrow f(x)$, We classically generate a training data set of the form
\begin{equation}
\label{eqn4}
T = \{ (\ket{x^{(i)}}, f(x^{(i)})  ) \}_{i=1}^{N}\,.
\end{equation}
In general, the amount of training data required could scale exponentially in the problem size (\ie number of qubits), since the data must be general enough to cover the space of possible inputs.

Recall that the parameters $\alv$ define a circuit $G_{\alv}$, which in turn defines a noisy quantum channel $\EC^{\rm Par}_{\alv}$. For this quantum channel, let $y^{(i)}_{\alv}$ denote the output of the circuit (\ie the expectation of the observable of interest) when the input is $\ket{x^{(i)}}$. Then we define the cost function as
\begin{equation}
\label{eq:obs_ext_cost}
C_{\OBE}(\alv) = \frac{1}{N} \sum_{i=1}^N (f(x^{(i)}) - y^{(i)}_{\alv} )^2\,.
\end{equation}
The cost quantifies the discrepancy between the desired output $f(x^{(i)})$ and the true output $y^{(i)}_{\alv}$, averaged over all training data points.

\subsubsection{State Preparation}

A second class of applications outlined in Fig.~\ref{fig:Applications} is state preparation. Here, the input is a quantum state or more generally a set of quantum states $\{\ket{x^{(i)}}\}_{i=1}^N$. The task is then to construct a circuit $U$ that prepares the output states $\{\ket{y^{(i)}} = U\ket{x^{(i)}}\}_{i=1}^N$ from these input states. In other words, one wishes to learn a unitary $U$ that accomplishes the desired state preparation task on the training data, $\{\ket{x^{(i)}}, \ket{y^{(i)}} \}_{i=1}^N$. Note that this is an under-constrained problem since in the state preparation application $N\ll 2^n$, where $U$ is an $n$-qubit unitary.  In this case, we use the following cost function:
\begin{align}
\label{eq:spcost}
    C_{\SP}(\alv) =  1 - \frac{1}{N}\sum_{i=1}^N F(\EC^{\rm Par}_{\alv}, \mathcal{U}, \ket{x^{(i)}}),
\end{align}
where $\UC(\cdot) \equiv U (\cdot) U\ad$. This is the infidelity between state prepared by $\EC^{\rm Par}_{\alv}$ and the target state $\ket{y^{(i)}}$, averaged over the training data points.
A typical scenario is when there is a single input and output state ($N=1$), as we will consider below in Section~\ref{sec:impl_stateprep}.

\subsubsection{Compilation}

Finally we consider the application of compiling a target unitary, $U$, into a set of native gates. The action of $U$ on all possible input quantum states must be reproduced. This is a more challenging task than constructing a state-preparation circuit, since one must consider the action on all states rather than just on one state or a small set of states.

Let $\UC(\cdot) \equiv U (\cdot) U\ad$ denote the quantum channel associated with $U$. Then we define the cost function for compiling as
\begin{align}
\label{eq:uccost}
    C_{\UCC}(\alv) = 1 - \Fbar(\EC^{\rm Par}_{\alv}, \UC) .
\end{align}
Note that this is analogous to Eq.~\eqref{eq:spcost} with the discrete average replaced by a continuous average (\ie integral with Haar measure). The average, $\bar{F}$, can be computed in various ways. Most elegantly, the average process fidelity is related to the entanglement fidelity $F_e$, via \cite{schumacher1996sending, horodecki1999general, nielsen2002simple}
\begin{align}
 \Fbar(\EC_{\alv}, \UC) = \frac{d F_e(\UC\ad \circ \EC^{\rm Par}_{\alv}) +1}{d+1}\,,
 \label{eq:avgfid_entfid}
\end{align}
where $F_e(\EC) = \mte{\phi}{\IC \ot \EC(\dya{\phi})} = F(\ket{\phi}\bra{\phi}, \mathcal{E}(\ket{\phi}\bra{\phi}))$, with $\ket{\phi} = \sum_j \ket{j}\ket{j}/\sqrt{d}$ being a maximally entangled state, and $d=2^n$ being the Hilbert-space dimension. 
Therefore, we can compute the compilation cost function by computing $F(\ket{\phi}\bra{\phi}, \mathcal{I}\otimes \mathcal{U}^\dagger \circ \mathcal{E}_{\alv}^{\rm Par}(\ket{\phi}\bra{\phi}))$. From the machine learning perspective, the training data set in this case just consists of a pair of states $\{\ket{\phi}, (\id \otimes U)\ket{\phi}\}$. However, this approach requires a computation in a doubled space of dimension $2^{2n}$. 

Alternative approaches to computing $\bar{F}$ that trade this greater memory complexity for greater time complexity (but can be easily parallelized) are (i) to approximate the Haar average with a sample average over a set of states that form a 2-design, or (ii) to use Nielsen's formula in terms of Pauli operators $\{\sigma_i\}_{i=1}^{d^2}$ ~\cite{nielsen2002simple}
\begin{align}
	\Fbar(\EC_{\alv}, \UC) = \frac{1}{d^2(d+1)}\Big( \sum_{i=1}^{d^2} \Tr(U\sigma_i U\ad \EC(\sigma_i) )+d^2 \Big). \nn
\end{align}
From the machine learning perspective, for (i), the training data set corresponds to the sampled 2-design and the the action of the ideal channel on these, $\{\ket{\phi_i}, U\ket{\phi_i}\}$, and for (ii) the training data set corresponds to the Pauli operators and the action of the ideal channel on these, $\{\sigma_i, U\sigma_i U\ad\}$.

\subsection{Machine learning algorithms} \label{sec:Optimization}

In this Section we describe machine learning methods that we use to find quantum circuits. The general idea behind our approach is the principle of task-oriented programming. The method should work with minimal information about the quantum algorithm, like the system size, number of ancilla qubits, measurement type or details about the target quantum computer on which the circuit will be run. With that minimal input, our method is supposed to find the best performing circuit that achieves the initially specified task.

The task is defined by the choice of training data and the form of a cost function. Here, the choice depends on particular application. The general rule is that the training data exemplifies the action of the algorithm that needs to be found. Generic example of a training data is given in Eq.~\eqref{eqn4}. There, the training set is generated by a function $f$ that encodes initially specified task. In Sec.~\ref{sec:impl_obs}, we use that framework to construct a training set for the quantum algorithm for computing state overlap. In this case, the function $f$ in Eq.~\eqref{eqn4} takes the form of a trace and the training data is defined as
\begin{equation} \label{eq:training_overlap}
T = \big\{
\big(
(\rho_i,\sigma_i) , \mathrm{Tr}(\rho_i \sigma_i)
\big)
\big\}_{i=1}^N \ .
\end{equation}

The size of the training data is expected to grow exponentially with the system size for many applications due to exponential size of the Hilbert space. Nevertheless, there are applications that do not require exponentially large training data sets. One such example is (possibly multiple) state preparation discussed in Sec.\ref{sec:impl_stateprep}. In the present work we used $N=15$ training data points in observable extraction example in Sec.~\ref{sec:impl_obs}. Other examples considered in the paper are trivial from the training data construction point of view. The choice of data in the training set affects the discovered algorithm, as we discuss in Sec.~\ref{sec:impl_obs}. In most applications one chooses the data to be as representative for a given task as possible. However, there is an interesting alternative to specialize the quantum algorithm by restricting the type of input data.

The choice of the cost function depends on the application. It is typically defined to measure the discrepancy between the action of the current algorithm under optimization and the expected action. The value of the cost function should reach zero if and only if the algorithm reproduces the specified task exactly. Section~\ref{sec:CostFunctions} gives details on the choice of the cost function for specific tasks considered in the paper.

We use regression machine learning algorithms to learn the relationship between inputs and outputs specified in the training data. In other words, we want to learn how to implement function $f$  in Eq.~\eqref{eqn4} (e.g. the trace in Eq.~\eqref{eq:training_overlap}) as a quantum circuit. It is done by minimizing the cost function over the space of quantum algorithms. This space is described by a set of parameters $\alv = (L,\vec{k},\vec{\theta})$, where $L$ is the depth of the circuit, $\vec{k}$ is a set of discrete parameters describing the layout of the algorithm and $\vec{\theta}$ are continuous parameters that span some of the quantum gates as detailed in Sec.~\ref{sec:GateSequence}.

 We stress that the way to include information about the noise model in our machine learning algorithm is to evaluate the cost function in a noisy simulator which implements that noise model. This includes creating a device model described with certain parallelization rules as described in Sec.~\ref{sec:NoiseModels}.

The optimization methods play an important role in our approach. The methods we describe here are general and can be applied to any type of the cost function. In particular, they are applicable to the cost functions associated with the applications discussed in Section~\ref{sec:CostFunctions}.

The space in which the optimization takes place is large and has a complicated form. In our method we are optimizing over circuits composed of gates taken from a particular alphabet. The circuit is described by two kinds of parameters, discrete and continuous. The discrete parameters $\vec{k}$ define the circuit's layout. That is, they specify what type of gate is acting on a given qubit, at a given time during the evaluation of the circuit. The continuous parameters $\vec{\theta}$ span all gates that contain at least one variational parameter. In the example of an alphabet derived from the IBM Q Ourense device in Eq.~\eqref{eq:ouresne_gs}, only $Z$ rotations contain a continuous parameter. 

The optimization is an iterative procedure in which every iteration is organized in two nested loops. In the inner one, the optimizer deals with continuous parameters with a fixed circuit layout $\vec{k}$. Changes to the structure of the circuit are introduced in the outer loop. The optimization over continuous parameters $\vec{\theta}$ is straightforward. Once the structure parameters $\vec{k}$ are fixed, the cost function depends on at most $L$ continuous parameters $\theta_i$. We use adaptive mesh based, gradient free, unconstrained (the cost function is invariant under $\theta_i \rightarrow \theta_i + 2\pi$) methods~\cite{abramson2009orthomads} to find a minimum of the cost function $C_{\vec{k}} = C_{\vec{k}}(\vec{\theta})$.

When the minimum $c$ of $C_{\vec{k}} = C_{\vec{k}}(\vec{\theta})$ is found, the optimizer switches to the outer loop and makes a change in the structural parameters $\vec{k}$. In this part of the procedure the optimizer is testing small, random updates to the structure of the circuit. Those updates include gate shuffling, gate removal as well as inserting new gates in the form of resolutions of identity (1-qubit and 2-qubit ones). This way, the number of gates $L$ in the circuit is variable and reaches an optimal (noise dependent) value during the optimization, see below for a more detailed discussion.
The circuit is also periodically compiled using simple, standard techniques. Here, we check for gate cancelations and simplification  that arise from commutation relations.
In principle, we could leave that task to machine learning as well but we find that doing it explicitly speeds up the learning process.
After new structural parameters $\vec{k'}$ are identified, the optimizer enters the inner loop and varies continuous parameters $\vec{\theta}$ to find a new minimum $c'$ of a cost function $C_{\vec{k'}} = C_{\vec{k'}}(\vec{\theta})$. Finally, the optimizer makes a decision whether or not the old circuit structure $\vec{k}$ should be replaced by the new one $\vec{k'}$. Here we follow the simulated annealing approach and accept the change if $c' < c$. The change is also accepted if $c' > c$ with probability exponentially decreasing in $c'-c$.

The above describes one iteration of the optimization algorithm. The iterations are repeated until convergence of the cost function is observed. The optimization is also restarted multiple times to detect possible local minima. 

Finally, let us mention an important feature of the optimization approach. As stated above, random structure updates done in the outer loop involve identity insertion and gate removal. Because the cost function is evaluated in the presence of noise, this procedure can sometimes lead to a larger value of the cost function (this is not possible with noiseless simulator). Thanks to that, the optimization algorithm automatically finds the optimal length $L$ of the circuit for a specified error model. Other machine learning approaches that are not noise-aware must be artificially biased towards short circuits. In contrast, our approach automatically finds a balance between deep, expressive but noisy circuits and shallow, less noisy ones.

\section{Noise model} \label{sec:nm}

We demonstrate NACL in the following sections using a fine-grained noise model derived from one- and two-qubit gate-set tomography (GST) \cite{blume-kohout_demonstration_2017, noauthor_pygsti._nodate,nielsen2020probing} experiments run on the five-qubit IBM Q Ourense superconducting qubit device. We emphasize that we are not claiming to capture the full behavior of this device; this cannot be done with just one- and two-qubit GST, and we need to make some assumptions about device behavior. The most important physical effects we are ignoring in this noise model are: (i) non-uniformity across the device, since we use one-qubit GST results on qubit 0 and two-qubit GST on the qubit pair 0-1 to infer process matrices for all qubits on the device, and (ii) since we do not characterize spectator qubits, we do not capture any crosstalk effects.

\begin{figure}[t!]
\includegraphics[width=0.3\columnwidth]{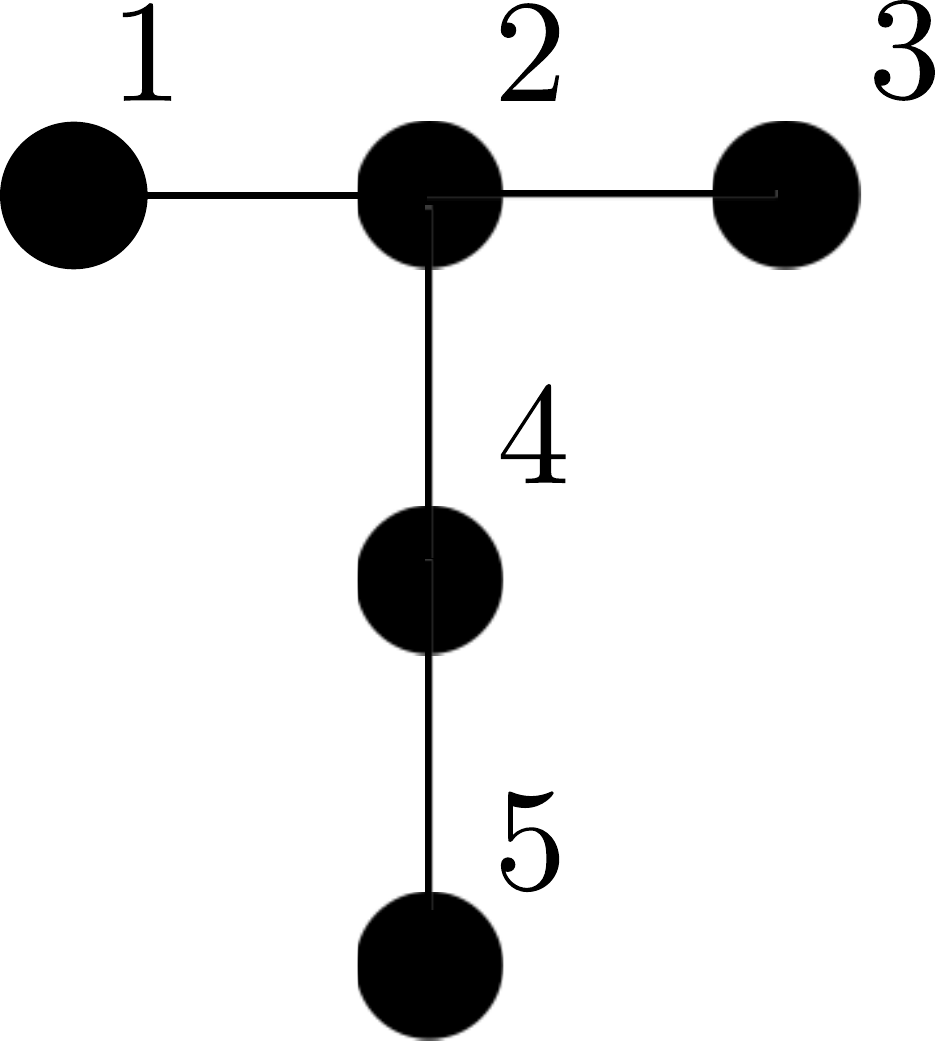}
\caption{Qubit layout and connectivity for device modeled in the noise model used to demonstrate NACL. This layout is inspired by the IBM Q Ourense device, and the lines indicate the qubits that can participate in CNOT coupling gates. 
\label{fig:nm}}
\end{figure}

One-qubit GST on qubit 0 of the Ourense device yields estimated one-qubit process matrices representing channels associated to the principal native gates on the device, $X(\pi/2)$ (or the ``pulse'' gate), and $I$, the single-qubit idle operation. The other single qubit gate used in this device is $Z(\theta)$, but this is performed virtually in software (through a phase shift of future single qubit gates) and so we assume it takes no time and is implemented perfectly. We also use the process matrices estimated by single-qubit GST for $\ket{0}$ state preparation and single-qubit measurement POVM elements for representing these operations. Then two-qubit GST on qubits 0 and 1 is used to extract a process matrix for the CNOT gate.  All the estimated process matrices and their figures of merit are presented in Appendix~\ref{sec:proc_mat}.

We assume that the layout and connectivity of the qubits are the same as for the IBM Q Ourense device, and these are outlined in Fig.~\ref{fig:nm}. This connectivity and the process matrices described above together define our \emph{device model}.

Note that we only performed GST on qubits 0 and~1 for simplicity, and assume that the resulting process matrices describe the same gates on other qubits also. This assumption could easily be relaxed at the expense of more GST experiments on all the qubits in the device. 

In the following sections, we will demonstrate NACL with a device model composed of the connectivity information for IBM Ouresnse and the above process matrices obtained through GST. All cost functions will be evaluated through simulation of circuits based on this device model.

\section{Implementation for Observable Extraction} 
\label{sec:impl_obs}

\begin{figure}[t!]
\includegraphics[width=\columnwidth]{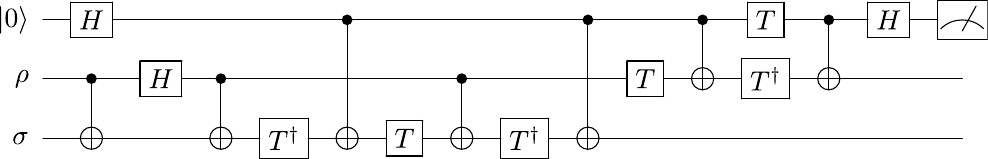}
\caption{The textbook SWAP-test based circuit for state overlap estimation when the input states ($\rho$, $\sigma$) are single qubit states. It is obtained by decomposing the SWAP operation into a standard universal gate set.
}
\label{fig:overlap_swapcircuit}
\end{figure}

The observable extraction application we focus on is state overlap estimation, where the task is to estimate the overlap between two input states $\rho$ and $\sigma$, \ie estimate ${\rm Tr}(\rho\sigma)$. The standard way to achieve this is to apply a controlled swap operation conditioned on an ancilla qubit, and then measure an expectation of an observable on the ancilla. We consider the case where $\rho$ and $\sigma$ are single qubit states, and decompose the textbook SWAP-based circuit for overlap estimation into a standard gate set in Fig. \ref{fig:overlap_swapcircuit}.

\begin{figure}
\includegraphics[width=\columnwidth]{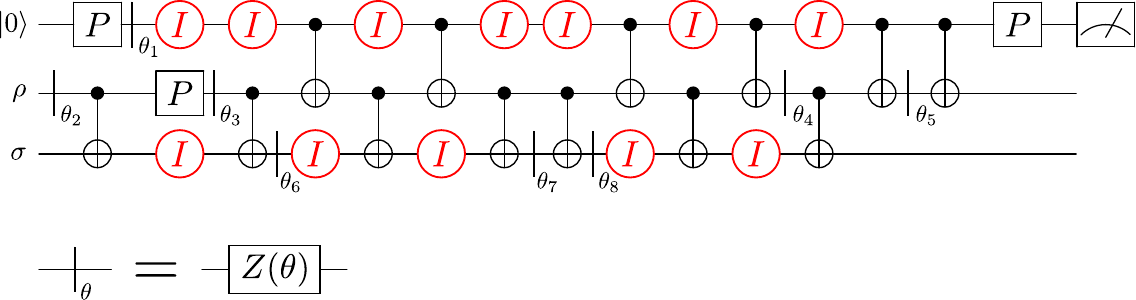}
\caption{The form of the textbook SWAP-test based overlap estimation circuit, shown in Fig. \ref{fig:overlap_swapcircuit}, when decomposed into the native gates in our device model. $P$ denotes the pulse gate, or $X(\pi/2)$ rotation, and $I$ is an idle timestep. The vertical lines denote $Z(\theta)$ rotations that are done virtually and therefore take no time. This notation helps visualize which gates can be performed in parallel. Values of $\theta_n$ are shown in Appendix~\ref{sec:thetas}.}
\label{fig:overlap_textbook_decomposed}
\end{figure}

For evaluation under the noise model, we first compile the textbook circuit in Fig.~\ref{fig:overlap_swapcircuit} into the native gate set composed of CNOT, $X(\pi/2)$ and $Z(\theta)$ rotations. Given the connectivity of the device, Fig.~\ref{fig:nm}, we map the input qubits to qubits 2 and 3, and the ancilla qubit to qubit 1. This is the most favorable mapping since in this case the minimal number (2) of CNOTs in Fig.~\ref{fig:overlap_swapcircuit} needs to be decomposed to account for the lack of device connectivity. There are other mappings that result in similar requirements for CNOT decomposition. We iterated over all of them and selected the decomposition that gives the smallest error (as measured by the value of the cost function evaluated in the presence of noise).

The decomposed circuit is shown in Fig.~\ref{fig:overlap_textbook_decomposed}. In this figure we show identity gates, or periods where a qubit is idle, in red. This circuit has been compressed and made as parallel as possible (using simplifications afforded by simple commutation relations and circuit identities), however, the remaining idle periods cannot be compressed away. We assume that $X(\pi/2)$ rotations (denoted $P$ in the figure) take the same amount of time as a CNOT for simplicity. 

\begin{figure}
\includegraphics[width=\columnwidth]{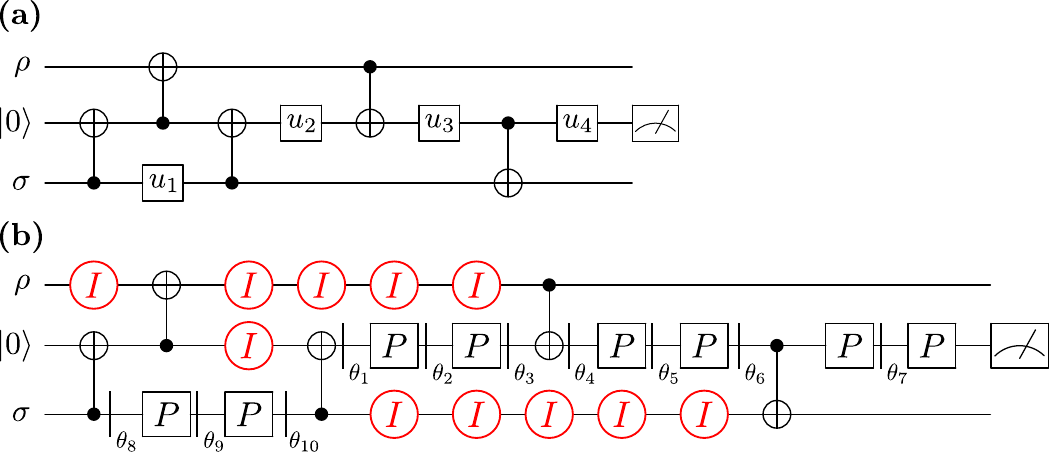}
\caption{\textbf{(a)} Machine learned circuit found without considering the noise model. \textbf{(b)} The circuit decomposed into the native gates in the device model. The notation is the same as in Fig. \ref{fig:overlap_textbook_decomposed}. Values of $\theta_n$ are shown in Appendix \ref{sec:thetas}.}
\label{fig:overlap_oldML}
\end{figure}

Next, we consider ML-based circuit implementations that \emph{do not} consider noise. Using techniques developed in \cite{cincio2018learning}, which attempt to find exact implementations that consist of as few gates as possible, we perform training without the noise model (but with the connectivity restrictions of the device). The training dataset size consists of 15 pairs of randomly generated single qubit states and their computed overlap. The resulting circuit for overlap estimation and its compiled version are shown in Fig.~\ref{fig:overlap_oldML}. In the absence of the noise model there is no penalty for the circuit to contain identity gates, and so the resulting circuit has a lot of them. 

\begin{figure}
\includegraphics[width=\columnwidth]{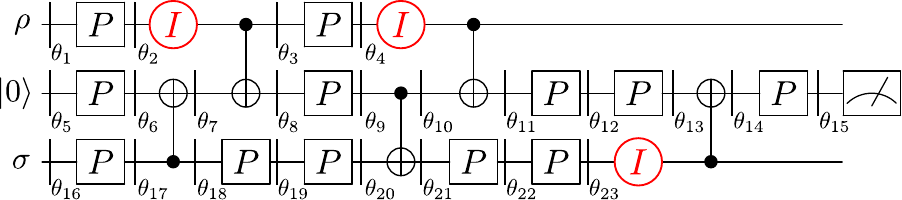}
\caption{Machine learned circuit found by NACL incorporating the noise model. The notation is the same as in Fig. \ref{fig:overlap_textbook_decomposed}. Values of $\theta_n$ are shown in the Appendix \ref{sec:thetas}.}
\label{fig:overlap_noisyML}
\end{figure}

Finally, we apply NACL to this problem and formulate the cost function using circuit simulation with the noise model described in Sec. \ref{sec:nm}. The training dataset size consists again of 15 pairs of randomly generated single qubit states and their computed overlap. The algorithm works directly with the native gate set, and so no subsequent decomposition is necessary. The circuit found by NACL is shown in Fig.~\ref{fig:overlap_noisyML}. 
Two features of the NACL circuit immediately stand out. First, since we have taken into account the noise associated with idling qubits, the circuit contains very few idles. Second, NACL makes interesting use of $Z(\theta)$ gates -- these are error free, take no time, and also increase the expressiveness of a circuit -- and consequently, NACL seems to maximize their use (especially compared the noise unaware ML circuit in Fig. \ref{fig:overlap_oldML}, which does not distinguish $Z(\theta)$ gates from other gates, and therefore does not use them more frequently). This liberal use of $Z(\theta)$ gates most likely also leads to the shorter depth circuit.
It should be stressed that these features are not built into the algorithm but result from the optimization and represent the best found balance between the number of gates and the noise induced by their action.

\begin{figure}
\includegraphics[width=\columnwidth]{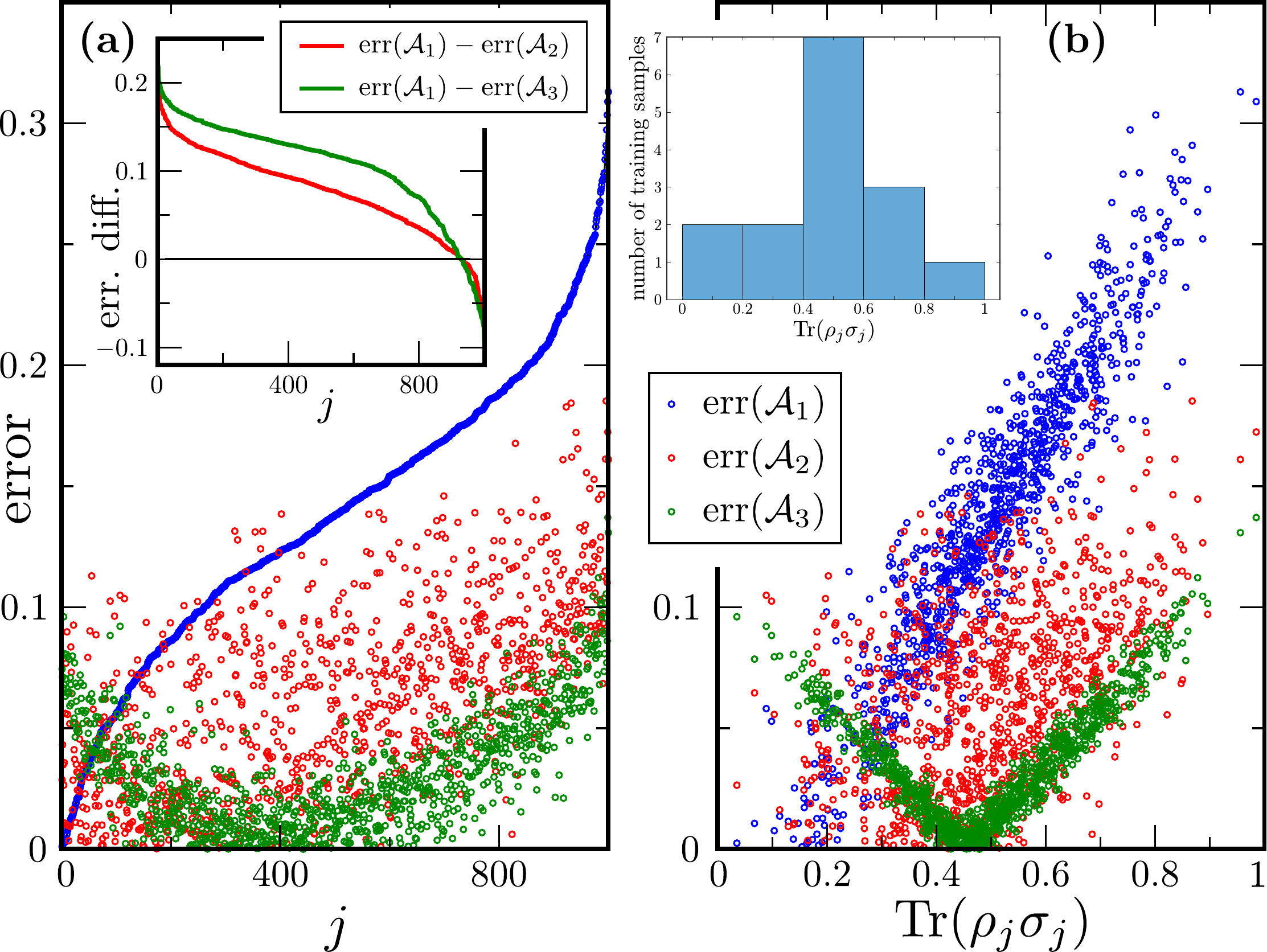}
\caption{\textbf{(a)} A comparison of error in computing state overlap (as quantified by Eq. \eqref{eq:err_obs}) for each of the validation samples for the three circuits: textbook ($\mathcal{A}_1$, blue), noise unaware ML ($\mathcal{A}_2$, red), and NACL ($\mathcal{A}_3$, green). The x-axis indexes pairs of states in the validation dataset. The inset shows differences in error. \textbf{(b)} Overlap estimation error for the three circuits as a function of the exact value of the overlap $\mathrm{Tr}(\rho_j \sigma_j)$. The inset shows a histogram of exact overlaps for the validation dataset.}
\label{fig:overlap_errors}
\end{figure}

In the following, we compare the performance of the three circuits described above. We generated a validation dataset -- 1000 pairs of new random one-qubit, mixed states $\{ \rho_j , \sigma_j \}$ -- and apply the three circuits to estimate the overlap between each pair (the circuits are simulated under the noise model). For simplicity, we label the textbook circuit (Fig. \ref{fig:overlap_textbook_decomposed}) $\mathcal{A}_1$, the noise unaware, standard ML circuit (Fig. \ref{fig:overlap_oldML}) $\mathcal{A}_2$, and the result of NACL (Fig. \ref{fig:overlap_noisyML}) $\mathcal{A}_3$. 
Fig.~\ref{fig:overlap_errors}(a) compares the errors of all three circuits, defined as the absolute value of the difference between the exact overlap $\mathrm{Tr}(\rho_j \sigma_j)$ and its estimate computed with the given circuit:
\begin{align}
	{\rm error}_{j,\mathcal{A}_i} = |{\rm Tr}(\rho_j\sigma_j) - \langle \sigma_z\rangle_{\mathcal{A}_i}|, 
	\label{eq:err_obs}
\end{align}
where $\langle \sigma_z\rangle_{\mathcal{A}_i}$ is the expectation value of the $\sigma_z$ operator on the measured qubit at the end of circuit $\mathcal{A}_i$.
The data is sorted such that the error of $\mathcal{A}_1$ is increasing with sample index, $j$. Fig.~\ref{fig:overlap_errors}(a) shows that noise-aware ML generated circuit gives the best overlap estimate for most of the state-pairs. 

The inset in Fig.~\ref{fig:overlap_errors}(a) shows the difference between the error of the textbook circuit and both ML circuits (these sets of data points are both independently ordered according to decreasing error difference). The ML circuit is better than the textbook one if the value shown in the inset is positive. We can see that this is indeed the case for over 90\% of cases, with the NACL circuit also outperforming the regular ML circuit in these cases. 
For further analysis, we look at the same data in Fig.~\ref{fig:overlap_errors}(b), but this time with the errors plotted against exact overlap of the 1000 samples in the validation dataset. This figure shows that the error of $\mathcal{A}_1$ generally decreases with the exact overlap. In addition, the error of $\mathcal{A}_3$ (NACL) shows non-monotonic behavior with exact overlap, achieving its minimum around exact overlap of 0.5 and increasing for larger and smaller overlaps. This behavior of NACL error can be explained by the specifics of training method and the type of the cost function that was used. NACL is trying to minimize average error (see Eq. \eqref{eq:obs_ext_cost}), and examining a histogram of overlaps in the training sample (inset in Fig. \ref{fig:overlap_errors} (b)) we see that these overlaps are concentrated between 0.4 and 0.5. Therefore,  NACL optimizes the average-case cost function by performing best on input state pairs that have overlaps around this value. An interesting observation is that there can be a correlation between the structure of a circuit and the overlaps it can best estimate. 

Finally, we can explain why the textbook circuit outperforms NACL in regions of low exact overlap as a combination of two factors: (i) as mentioned above, NACL minimizes average error, and the contribution to this from training samples with small overlap is small; hence it sacrifices performance on small overlap states to get better performance on states with larger overlap; (ii) the other factor that results in the textbook circuit performing well for small exact overlap samples is accidental; namely, that  the overlap is estimated by measuring $\langle \sigma_z \rangle$ on the ancilla, and this quantity tends to zero with circuit length (since the stochastic noise in the gates dampens this polarization). The output of $\mathcal{A}_1$ is small due to noise, and thus is \emph{accidentally} close to the correct answer for small overlap states.

We note that the uneven behavior of NACL with exact overlap of input states can be easily modified by (i) modifying the training dataset to have uniformly distributed overlaps, and (ii) modifying the cost function to be a worst-case measure of performance instead of average-case and/or a function of relative error as opposed to absolute error with the exact overlap.

\section{Implementation for state preparation}
\label{sec:impl_stateprep}
For the state preparation application, we will focus on preparing W-states of $n$ qubits:
\begin{align}
	\ket{W_n} = \frac{1}{\sqrt{n}}\sum_{i=1}^n \ket{i},
\end{align} 
where $\ket{i}$ is the state where qubit $i$ is $\ket{1}$ and all other qubits are in state $\ket{0}$. W-states are multipartite entangled states that are robust against loss and can be used for multipartite cryptographic protocols and for teleportation \cite{joo_quantum_2003}. As far as we are aware, the circuits generated in Cruz \etal \cite{cruz2019efficient} are the most efficient circuits for W-state generation, and we will use these circuits as our base-case ``textbook'' circuits to compare against.

In the following we will study the prepartion of W-states for $n=4,5$.

\subsection{4 qubit W-state preparation}

\begin{figure}[t!]
\includegraphics[width=\columnwidth]{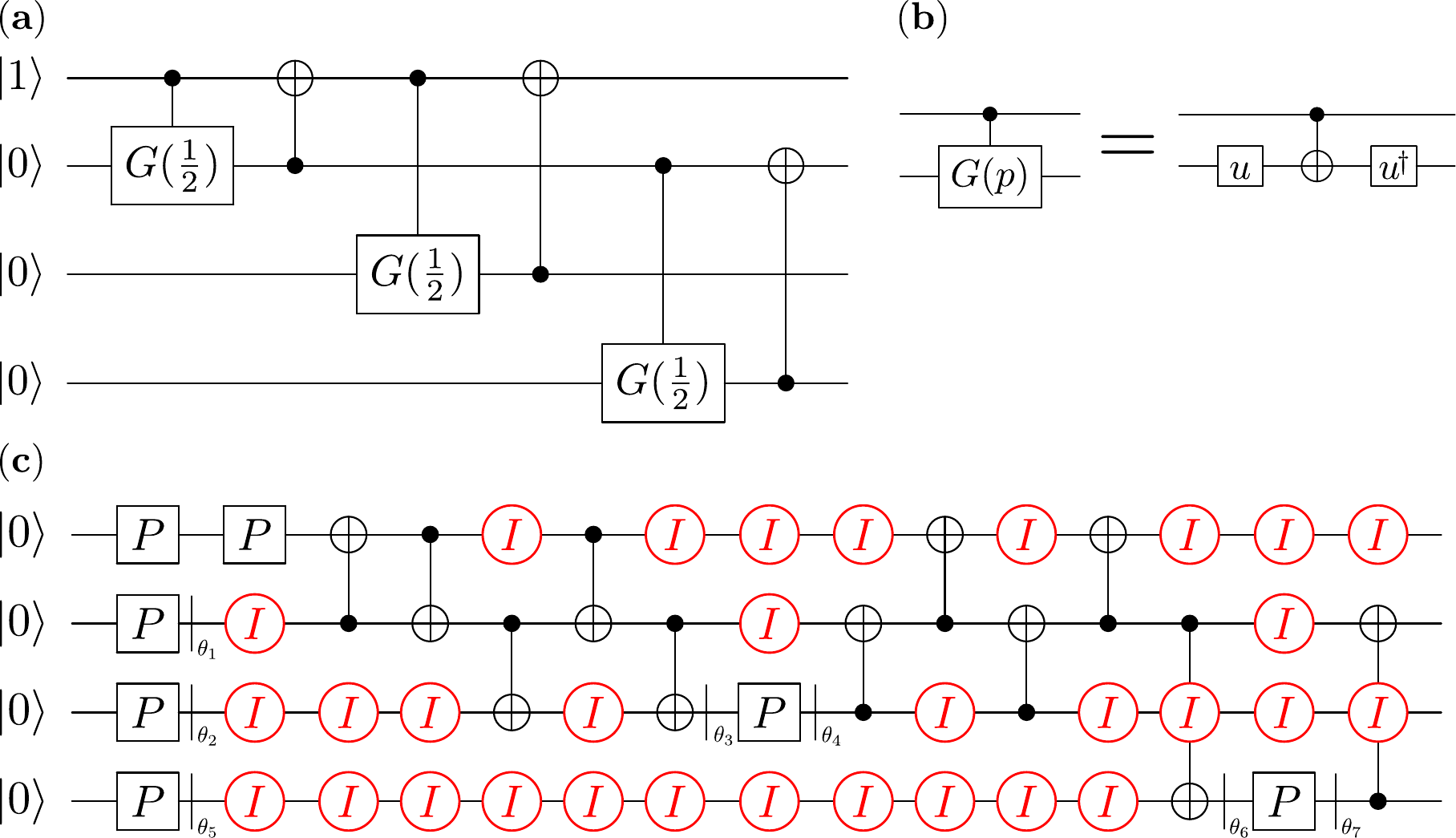}
\caption{\textbf{(a)} Textbook circuit for preparing $\ket{W_4}$. \textbf{(b)} Decomposition of control-$G(p)$ gate into CNOT and one-qubit gates $u$ and $u^\dagger$, where $u=e^{-iY\alpha}$ and  $\alpha =  \arcsin(\sqrt{p})/2$. \textbf{(c)} Compilation of the textbook circuit shown in (a) into first four qubits of the device model in Fig. \ref{fig:nm}. The notation is the same as in Fig. \ref{fig:overlap_textbook_decomposed}. The values of angles $\theta_j$ are given in Appendix~\ref{sec:thetas}.}
\label{fig:W4_textbook}
\end{figure}

The textbook circuit for preparing $\ket{W_4}$ is shown in Fig.~\ref{fig:W4_textbook}(a). It was obtained by following the general procedure given in \cite{cruz2019efficient}. This circuit will be applied to the first four qubits in the device shown in Fig. \ref{fig:nm}. 
The performance of the textbook circuit and the NACL circuit will depend on the subset of qubits on which we are preparing the state.
However, in realistic situations, one will not be given that freedom since the state preparation is usually only one step in a larger quantum circuit, which imposes constraints on the choice of qubits. We select qubits 1-4 to show how NACL can optimize circuits on devices with restricted connectivity.

The one-qubit gate, depicted as $G(p)$ in Fig.~\ref{fig:W4_textbook}(a), is defined as follows:
\begin{equation}
    G(p) = \left( \begin{matrix}
    \sqrt{p} & \sqrt{1-p} \\
    \sqrt{1-p} & - \sqrt{p}
    \end{matrix} \right) \ .
\end{equation}
Note that this is a slightly different definition than the one given in Ref. \cite{cruz2019efficient}. The above definition of $G(p)$ leads to the same state that is prepared with the circuit shown in Fig.~\ref{fig:W4_textbook}(a) but allows for more efficient decomposition of control-$G(p)$ into CNOTs and one-qubit gates. 

The circuit shown in Fig.~\ref{fig:W4_textbook}(a) must be compiled into the native gate set in the device model. The $W$ state is invariant under permutation of qubits, and so one can relabel the qubits in the circuit shown in Fig.~\ref{fig:W4_textbook}(a) if this is advantageous for compilation. To find the optimal compilation of the textbook circuit we checked all possible permutations of qubits. All permutations lead to a compilation in which at least two CNOTs are not compatible with device connectivity and need to be decomposed further. We evaluated each permutation by simulation (with the noise model) under the corresponding compiled circuit and computing the fidelity of the output with the exact $\ket{W_4}$ state. The permutation that gives the highest fidelity is simply $[1,2,3,4]$ (there are however other permutations that lead to the same fidelity), and the corresponding compiled circuit is shown in Fig.~\ref{fig:W4_textbook}(b). We found that this textbook circuit produces $\ket{W_4}$ with fidelity $0.671$ under the noise model.

\begin{figure}
\includegraphics[width=\columnwidth]{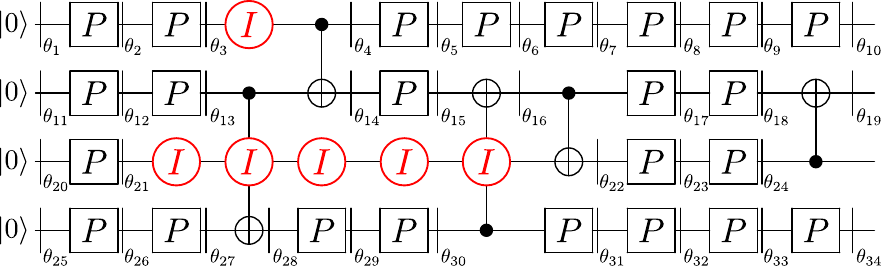}
\caption{Circuit that prepares $\ket{W_4}$ found by NACL. The notation is the same as in Fig. \ref{fig:overlap_textbook_decomposed}. Angles $\theta_j$ are specified in Appendix \ref{sec:thetas}.}
\label{fig:W4_ML}
\end{figure}

The circuit produced by NACL for preparing $\ket{W_4}$ is shown in Fig.~\ref{fig:W4_ML}. Since the task here is to prepare one state from one other state, the training dataset and validation dataset are the same, and just consist of one pair $\{ \ket{0}^{\otimes 4}, \ket{W_4} \}$; the first element is the input state and the second is the ideal output state. This NACL circuit outputs a state under the noise model with a fidelity of $0.8894$ to the exact state. This is a reduction in error (as measured by $1-F$, where $F$ is fidelity) by a factor of $3$ as compared with the best known textbook circuit.

Careful inspection of the circuit in Fig.~\ref{fig:W4_ML} reveals an interesting feature. In certain circumstances, it is more beneficial (from the point of minimizing the cost function; infidelity in this case) to have a long sequence of gates that are not compiled into an equivalent transformation with a shorter sequence. An example is the final 13 gates (including $Z(\theta)$ gates in this count) applied to qubit 1. It is possible to implement the resulting transformation with a shorter sequence of gates, but doing so would mean that the qubit sits idle for the remaining time while the operations on the other qubits complete. Apparently this incurs a greater cost than the longer sequence (the pulse gates are fairly high quality gates for this device and in fact, have a smaller infidelity than the idle operations, see Appendix \ref{sec:proc_mat}). We thus observe a feature that resembles dynamical decoupling or a dynamically corrected gate for this final transformation of qubit 1. We have reasonable confidence that this feature is not a numerical artifact or local optimum because we also independently optimized just that subcircuit (\ie keep the rest of the circuit fixed and optimized just the last six clock cycles of qubit 1 under the same cost function that evaluates the error on the 4-qubit output state), and could not find a better sequence. Note that this feature is ``emergent''. Dynamical gate correction techniques were not coded in the search algorithm and yet NACL effectively used them in the optimized solution. It a way, those techniques were ``discovered'' via cost optimization. We also point out that this feature of preferring longer sequences to idles is not general -- one cannot replace every sequence of idles with a sequence of pulses and $Z(\theta)$ rotations and lower the error. For example, qubit 3 sits idle over five clock cycles and this achieves the minimum cost function even when we attempt to re-optimize just that sub-sequence of gates. This feature demonstrates the ability of NACL to find circuit implementations that optimize performance in highly non-trivial ways that incorporate an interplay between the computational task (encoded in the cost function) and the device model.

\subsection{5 qubit W-state preparation}

\begin{figure}[t]
\includegraphics[width=\columnwidth]{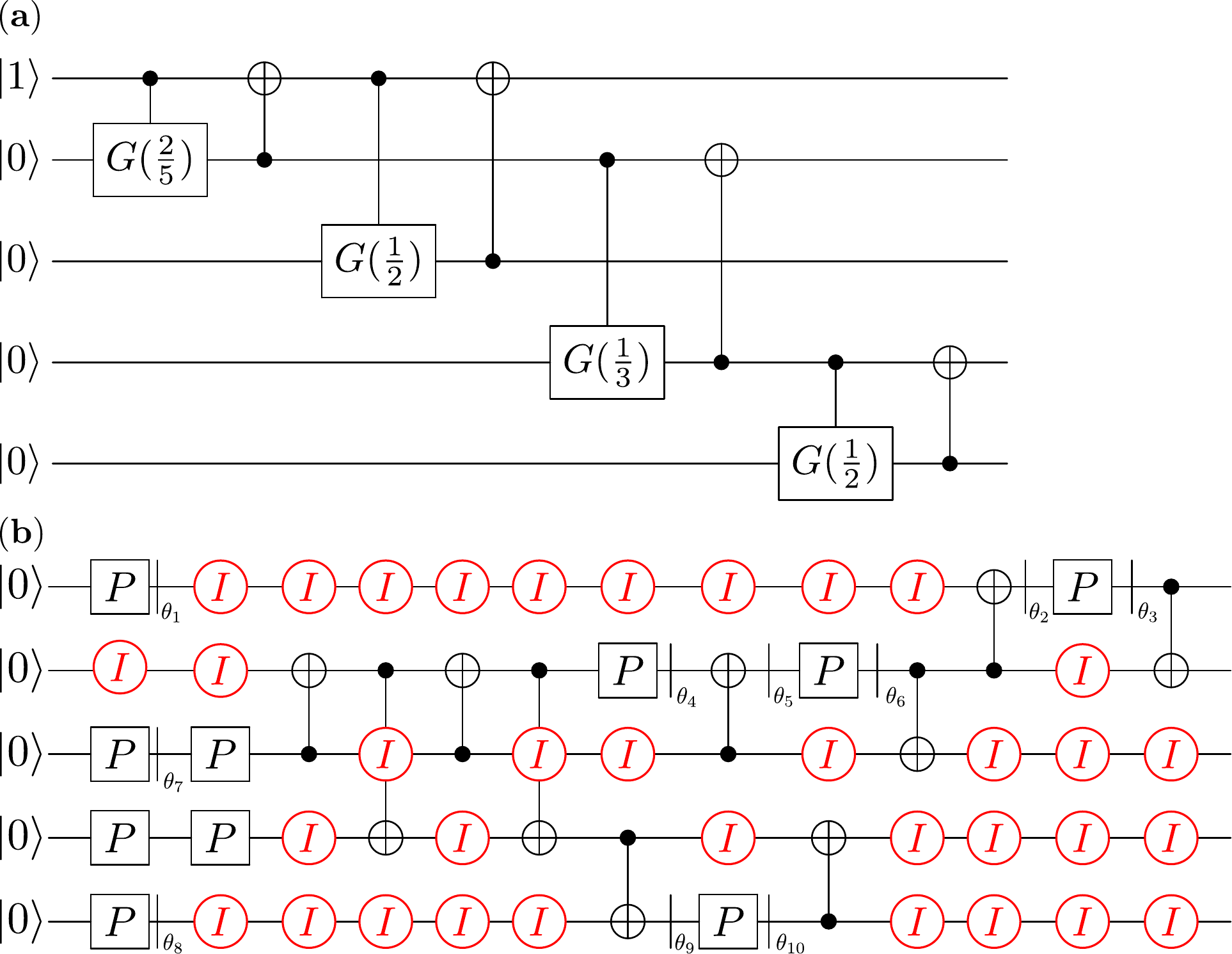}
\caption{\textbf{(a)} Circuit for preparing $\ket{W_5}$ obtained by following the construction given in \cite{cruz2019efficient}. The first controlled $G(p)$ gate can be simplified, as the first qubit is initialized in $\ket{1}$. This allows for a shorter compilation. \textbf{(b)} Its best compiled version achieved by a proper permutation of qubits. The notation is the same as in Fig. \ref{fig:overlap_textbook_decomposed}. Angles $\theta_j$ are given in Appendix \ref{sec:thetas}.}
\label{fig:W5_textbook}
\end{figure}

We also study the preparation of $\ket{W_5}$ since this task requires the use of all qubits on the device in Fig. \ref{fig:nm}. Again, we follow the prescription in Cruz \etal \cite{cruz2019efficient} to arrive at the best textbook circuit for preparing $\ket{W_5}$ in Fig.~\ref{fig:W5_textbook}. The compilation of this circuit onto the device under study is not trivial since we can arbitrarily permute the qubits. Every permutation will result in a potentially different decomposition of CNOTs, given the constrained connectivity of the device. We checked all 120 qubit permutations and found that the circuit compilation shown in Fig.~\ref{fig:W5_textbook}(b) gives the smallest value of the cost function when evaluated under the noise model. This optimal permutation was found to be $[4,3,5,2,1]$. Under this permutation, only one CNOT (the second gate from the left in Fig.~\ref{fig:W5_textbook}(a)) needs to be decomposed due to the lack of connectivity. The circuit in Fig.~\ref{fig:W5_textbook}(b) achieves the fidelity of $0.675$.

\begin{figure}[t]
\includegraphics[width=\columnwidth]{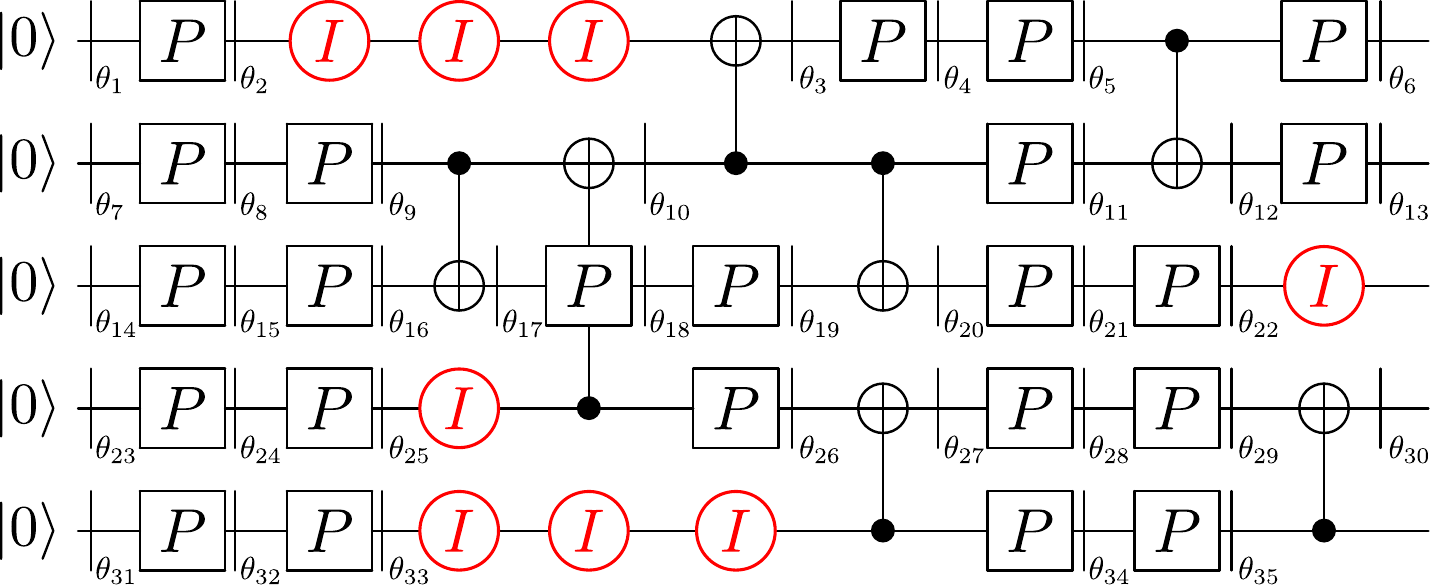}
\caption{The circuit that approximates preparation of $\ket{W_5}$ found by NACL. The notation is the same as in Fig. \ref{fig:overlap_textbook_decomposed}. Angles $\theta_j$ are given in Appendix \ref{sec:thetas}.}
\label{fig:W5_ML}
\end{figure}

NACL found the circuit presented in Fig.~\ref{fig:W5_ML} for $\ket{W_5}$ state preparation. Again, NACL finds a circuit that is much more compact than the textbook one. It uses fewer CNOTs, requires less idling of qubits, and uses the error-free $Z(\theta)$ gates liberally. The circuit produces an output state with fidelity of $F = 0.837$ with the ideal $\ket{W_5}$ state. That is, the error (as measured by $1-F$) is reduced by a factor of $2$ as compared to the textbook circuit.

\section{Implementation for circuit compilation}
\label{sec:impl_comp}
For the circuit compilation application we consider the problem of compiling the quantum Fourier transform (QFT), which is a paradigmatic building block that is used in many quantum algorithms \cite{nielsen_quantum_2010}. In the following we will consider implementing a three-qubit QFT. 

\begin{figure}
\includegraphics[width=\columnwidth]{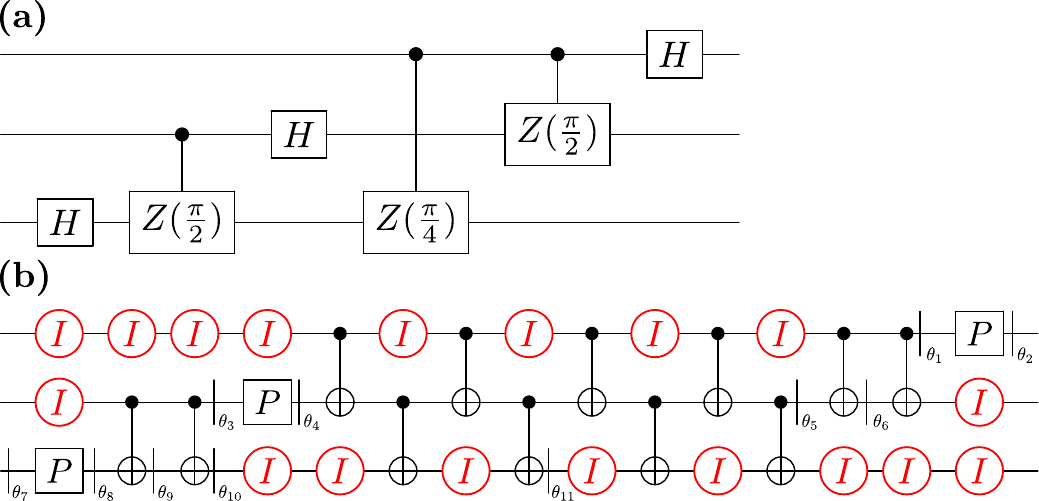}
\caption{\textbf{(a)} A textbook circuit for performing QFT on three qubits. \textbf{(b)} A compilation of the circuit in (a) into the native gate set in the device model we are simulating. The compilation has to take into account that qubit 1 and 3 are not directly connected. Angles $\theta_j$ are specified in Appendix ~\ref{sec:thetas}.}
\label{fig:QFT_textbook}
\end{figure}

A textbook circuit for implementing a QFT on three qubits is shown in Fig.~\ref{fig:QFT_textbook}(a). We will consider implementing this on qubits 1, 2 and 3 in the device shown in Fig. \ref{fig:nm}. We first need to decompose the controlled $Z(\theta)$ rotations. Every controlled $Z(\theta)$ is decomposed using two CNOTs \cite{cross2017open}. This decomposition leads to two CNOTs between qubits 1 and 3. Since these qubits are not directly connected, these CNOTs need to be decomposed further. The result of this compilation procedure is shown in Fig.~\ref{fig:QFT_textbook}(b). This compilation leads to a very sparse circuit with many (incompressible) idle gates, which has negative impact on the quality of the final result.

\begin{figure}
\includegraphics[width=\columnwidth]{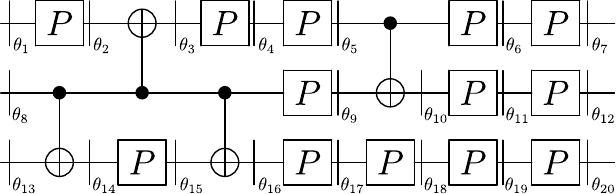}
\caption{Circuit performing QFT found by NACL. The notation is the same as in Fig. \ref{fig:overlap_textbook_decomposed}. Angles $\theta_j$ are given in Appendix \ref{sec:thetas}.}
\label{fig:QFT_NCL}
\end{figure}

The circuit constructed via NACL is shown in Fig.~\ref{fig:QFT_NCL}. We used NACL with the cost function defined in Eq.~\eqref{eq:uccost} with the average process fidelity computed via Eq.~\eqref{eq:avgfid_entfid}. The circuit has shorter depth than the compiled textbook circuit, and does not contain a single idle gate (as compared with 18 for the textbook circuit). It also contains more error-free $Z(\theta)$ rotations enhancing the expressiveness of the circuit.

\begin{figure}[t!]
\includegraphics[width=\columnwidth]{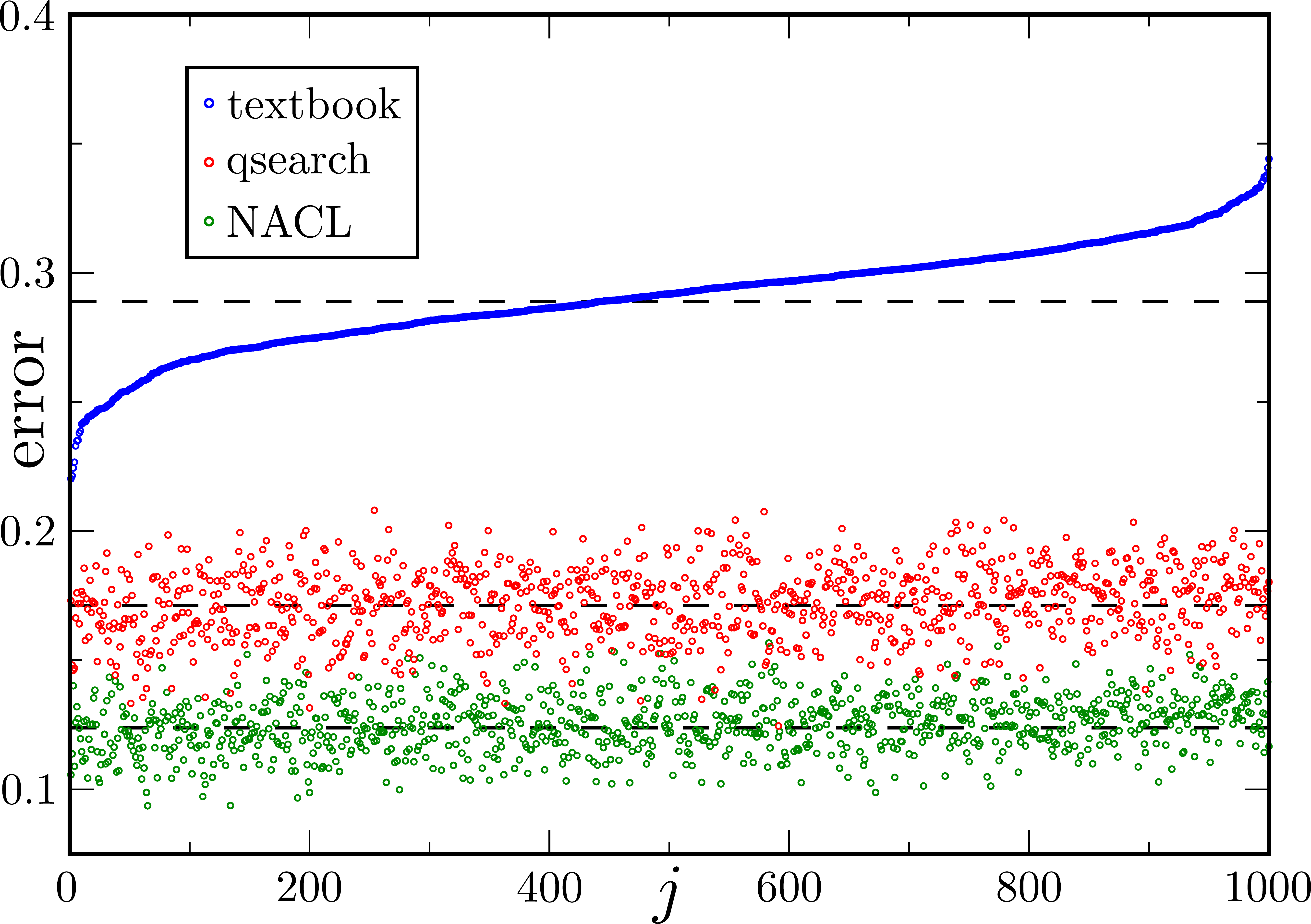}
\caption{Performance of textbook, qsearch-generated and NACL-generated circuits to evaluate QFT. The figure shows error (as defined in the text) for 1000 randomly generated pure states. NACL-generated circuit performs much better than the textbook one on all considered states. The circuit found by NACL has also a lower error than the one generated by qsearch. }
\label{fig:QFT_errors}
\end{figure}

To compare the performance of the two compiled circuits for QFT, we select 1000 random pure states $\ket{\Psi_j}$ and evaluate each circuit on those states. The error metric we use is the infidelity between the ideal QFT output and the circuit output; $1 - \mathrm{Tr}(\rho_j \ket{\Psi^\mathrm{ex}_j} \bra{\Psi^\mathrm{ex}_j} )$, where $\ket{\Psi_j^\textrm{ex}}$ is the result of the exact evaluation of QFT on $\ket{\Psi_j}$. Our results are summarized in Fig.~\ref{fig:QFT_errors}.
We also compare our results with qsearch~\cite{davis2020towards}, recently proposed technique for circuit compilation. Qsearch is typically used as an exact compilation method but it can be used with a finite precision $\delta$. Qsearch results shown in Fig.~\ref{fig:QFT_errors} are obtained by running it with various values of $\delta$ and selecting the circuit that gives the smallest error as defined by Eq.~\eqref{eq:avgfid_entfid}.
For easier comparison, the states $\ket{\Psi_j}$ were ordered such that the error of the textbook circuit (represented by the blue line) increases with the state index $j$. The NACL-generated circuit performed better than the textbook one on all considered states. It also outperform the circuit found by qsearch even after minimizing over compilation precision $\delta$. Since the validation dataset is composed of random pure input states, the average infidelity (over these input states) is related to the entanglement infidelity of the channel defined by the noisy circuit (see Eq.~\eqref{eq:avgfid_entfid}), which is an input-state independent measure of the quality of a channel (or circuit implementation).  We use this relation to validate our error metric defined over randomly sampled input states. In Fig.~\ref{fig:QFT_errors} the dotted lines show $1-(dF_e(\mathcal{U}^\dagger \circ \EC)+1)/(d+1)$, where $d=8$, $\mathcal{U}$ is the channel corresponding to the ideal circuit implementation, $\EC$ is the channel corresponding to the noisy circuit implementation, and $F_e$ is the entangled fidelity defined in Sec.~\ref{sec:CostFunctions}. These lines correspond well to the sample averages of our infidelity error metric. We find that NACL reduced the average infidelity of a textbook circuit from $0.289$ to $0.124$, that is, by 57\%. NACL also reduced the error by a factor of $1.4$ as compared to qsearch. Another observation is that the performance of the textbook circuit varies more significantly with input state than for the NACL-generated circuit.

\section{Discussion and Conclusions}
\label{sec:disc}

We have introduced the framework of noise-aware circuit learning (NACL), whereby the circuit implementation of a quantum algorithm is formulated by machine learning and optimization based on a cost-function that captures the goal of the algorithm and a device model that captures the connectivity and noise in the device that executes the circuit. We have shown that this framework can be applied to all of the common tasks in quantum computing -- observable (or mean-value) extraction, state preparation, and circuit compilation -- and demonstrated through examples the types of performance improvements that can be obtained through NACL. For the examples considered here, NACL produces reductions in error rates (suitably defined for the different tasks) by factors of 2 to 3, when compared to textbook circuits for the same tasks.

In general, NACL produces shorter depth circuits that minimize the impact of stochastic noise sources. However, as demonstrated through the examples considered here, NACL can automatically derive known noise-suppression concepts such as dynamical decoupling and apply these in contexts where they are useful (as defined by the cost function). It also naturally outputs circuits that incorporate commonsense strategies such as minimizing the number of noisy idle gates and maximizing the use of ideal gates, such as error-free $Z(\theta)$ rotations. NACL can incorporate much more fine-grained information about the device than other circuit compilation techniques -- \eg in the demonstrations presented here we have used process matrices derived from gate set tomography of real hardware to approximately model noise on this device. Such process matrices can capture effects ignored by effective noise models, such as coherent noise and non-unital processes such as relaxation. 

We note that we have also executed NACL with an error model derived from trapped-ion physics (see Appendix \ref{sec:eff_nm} for details), to validate that the technique can be used with a variety of noise model specifications. The results are very similar to those presented above, although there are some simplifications due to an assumption of full connectivity in the device (which is realistic for small trapped-ion platforms).

The noise models currently compatible with NACL do not include crosstalk effects. Although these can be incorporated for small devices using the approach outlined in this paper, incorporating crosstalk in a scalable manner is complicated. The heart of the issue is how to model crosstalk in a scalable manner \cite{sarovar_detecting_2019}.
In the presence of crosstalk, the natural description of operations on a quantum computer is not in terms of gates, but in terms of \emph{layers}, which capture what is done to each qubit in the device in a given clock cycle. This is because the precise operation performed on a qubit could, in principle, depend on what is performed on any other qubit in the device. Therefore, the first extension of NACL required to capture crosstalk is to optimize circuits in terms of layers as opposed to gate sequences. Moreover, one has to also consider whether it is realistic to develop quantum channels representing noisy implementation of any circuit layer. Firstly, there are an exponential (in $n$, the number of qubits) number of possible layers to characterize, and secondly, one needs to perform $n$-qubit process tomography in order to get quantum channels for each layer. This last task is obviously impossible for large $n$, and therefore one has to develop more approximate techniques to describe noisy implementations of layers. One approach around these issues is to patch together quantum channels derived from one-, two-, and three-qubit tomography to get an approximate description of a circuit layer, similar to what is demonstrated in Govia \emph{et al.} \cite{govia_bootstrapping_2020}. This would model a physically important subclass types of crosstalk errors \cite{sarovar_detecting_2019}. Another approach is to forego full tomographic information about error processes and instead use effective noise models or error rates that contain information about crosstalk, \eg see Ref. \cite{murali_software_2020} for an example of how such partial information can be used to model crosstalk errors. Of course, one is trading off NACL prediction accuracy when approximate noise models are used. Future work will look at incorporating these more complex noise effects into the NACL circuit learning framework. 

An important issue to consider is how to scale NACL to develop noise-resilient circuits for larger devices. The complexity of circuit simulation under a noise model and the complexity of optimization over the circuit parameters increase exponentially with number of qubits. 
As a consequence, the current NACL approach could be used as-is to optimize circuits on up to about 8-10 qubits. With code optimization and parallelization this could be extended to circuits on 12-14 qubits. Such machine learned noise resilient circuits could be useful for increasing the performance of small modular elements of larger circuit applications; \eg magic state distillation circuits.
However, we can also outline a strategy for extending NACL beyond such use-cases. The strategy applies when one is already given a circuit compilation for a computational task. Perhaps this is a compilation derived using theoretical decompositions or some other efficient method. Then one can sample a subcircuit from this circuit. This subcircuit defines an ideal unitary and one can use NACL to find best approximations to this unitary under the given device model. This sampling can be repeated for multiple subcircuits. However, note that this strategy does not guarantee any optimality properties for the circuit derived from combining these individually optimized subcircuits. Studying the potential of this strategy for scaling up the NACL framework is left as future work. 

Related to scalability is the connection between NACL and variational quantum algorithms (VQAs). An alternative to evaluating the NACL cost functions in Sec.~\ref{sec:CostFunctions} by simulating a parameterized quantum circuit on a classical computer is to evaluate them by executing the parameterized circuits on quantum hardware directly in the spirit of VQAs. In addition to the obvious advantage of scalability, this hardware-enabled approach has the advantage of capturing the noise model exactly (and does not require any noise modeling). However, for certain applications (\eg compiling and state preparation~\cite{Khatri2019quantumassisted, jones2018quantum}), the NACL cost function require comparing against the \emph{ideal} target circuit outputs. In a VQA setting, any preparation of the targets would also be noisy, and therefore one cannot exactly evaluate the required cost functions. Whether it is possible to sufficiently approximate the cost functions with noisy hardware is an open problem \cite{sharma2020noise}, and if this were possible, it would make hardware-enabled NACL realistic.

We note that NACL typically outputs approximations of the task that is specified. This is because of two reasons: (i) in the presence of typical noise models the best one can do is approximate an ideal unitary map, and (ii) NACL provides no guarantee of finding global minima of the cost functions, which are typically extremely nonlinear. Therefore, even if the noise model is benign enough that the global minimum/minima correspond to ideal implementations of the target unitary, NACL will most likely find a local minima that is an approximation of the target unitary. However, as empirically demonstrated in this paper  NACL output is often far superior to textbook derived circuits, or even circuits optimized using other compilation techniques.

Modern optimization and machine learning methods will be critical for deriving computational use from near-term quantum devices. Motivated by this, we have developed the NACL framework as a way to utilize detailed noise characterization information to build noise-resilient circuits for near-term quantum computing applications, and we outlined promising directions for extending this framework. Our NACL method can be combined with (and hence is complementary to) other approaches to error mitigation that have been recently proposed~\cite{temme2017error,kandala2019error,czarnik2020error,strikis2020learning,zlokapa2020deep}. Hence, NACL is a novel primitive that will play an important role in the quest for quantum advantage.

\begin{acknowledgements}
The authors would like to thank Tim Proctor and Andrew Baczewski for useful comments on a draft of this work.

Research presented in this article was supported by the Laboratory Directed Research and Development program of Los Alamos National Laboratory under project number 20180628ECR for the noise-free machine learning approach and project number 20190065DR for the machine learning approach in the presence of noise. PJC also acknowledges support from the LANL ASC Beyond Moore's Law project. This work was also supported by the U.S. Department of Energy, Office of Science, Office of Advanced Scientific Computing Research, under the Quantum Computing Application Teams (QCAT) program. Sandia National Laboratories is a multimission laboratory managed and operated by National Technology and Engineering Solutions of Sandia, LLC, a wholly owned subsidiary of Honeywell International, Inc., for the U.S. Department of Energy's National Nuclear Security Administration under contract DE-NA0003525. This paper describes objective technical results and analysis. Any subjective views or opinions that might be expressed in the paper do not necessarily represent the views of the U.S. Department of Energy or the United States Government.

\end{acknowledgements}
\bibliography{q.bib}

\clearpage
\widetext
\appendix

\section{Numerical values of rotation angles}
\label{sec:thetas}
In Table \ref{tab:angles} we list the angles $\theta_n$ that define the $Z(\theta)$ gates in all the circuits presented in the main text.

\begin{table}[hb!]
\centering
\begin{tabular}{|c|c|c|c|c|c|c|c|c|c|}
\cline{1-10}
$n$ & $\theta_n$ in                                                      & $\theta_n$ in                                          & $\theta_n$ in                                         & $\theta_n$ in                                     & $\theta_n$ in                               & $\theta_n$ in                                     & $\theta_n$ in                               & $\theta_n$ in                                         & $\theta_n$ in                                 \\
    & Fig. \ref{fig:overlap_textbook_decomposed} & Fig.\ref{fig:overlap_oldML}(b) & Fig. \ref{fig:overlap_noisyML} & Fig. \ref{fig:W4_textbook}(c) & Fig. \ref{fig:W4_ML} & Fig. \ref{fig:W5_textbook}(b) & Fig. \ref{fig:W5_ML} & Fig. \ref{fig:QFT_textbook}(b) & Fig. \ref{fig:QFT_NCL} \\ \cline{1-10} 
1 &      3.926991 & 4.729459 & 1.249703 & 1.570796 & 6.261032 & 0.785398 & 0.000110 & 2.748894 & 5.528124 \\ 
     2 & 1.570796 & 2.356455 & 2.410073 & 0.785398 & 1.581678 & 2.356194 & 3.107730 & 1.570796 & 1.603007 \\ 
     3 & 1.570796 & 6.271231 & 5.503714 & 2.356194 & 0.618247 & 3.141593 & 0.775046 & 2.356194 & 3.180240 \\ 
     4 & 0.785398 & 0.012770 & 0.172928 & 3.141593 & 6.210505 & 0.615480 & 1.385048 & 1.570796 & 1.588121 \\ 
     5 & 5.497787 & 5.497958 & 0.106332 & 0.785398 & 3.155136 & 2.526113 & 3.218745 & 0.785398 & 5.982926 \\ 
     6 & 5.497787 & 0.017911 & 0.022432 & 2.356194 & 3.088771 & 3.141593 & 6.184576 & 5.497787 & 1.517815 \\ 
     7 & 0.785398 & 0.785692 & 1.621729 & 3.141593 & 3.127992 & 1.369438 & 0.000214 & 1.570796 & 3.174252 \\ 
     8 & 5.497787 & 4.713347 & 6.267293 &          & 2.708279 & 0.785398 & 0.725692 & 2.356194 & 2.380614 \\ 
     9 &          & 2.355849 & 3.672289 &          & 1.477670 & 2.356194 & 0.895856 & 5.497787 & 3.909589 \\ 
     10 &         & 4.711034 & 0.132619 &          & 2.327048 & 3.141593 & 0.149143 & 0.392699 & 6.271246 \\ 
     11 &         &          & 5.697289 &          & 0.012648 &  & 2.289056 & 5.890486         & 0.006795 \\ 
     12 &         &          & 3.141953 &          & 0.876330 &  & 3.142930 &                  & 3.903039 \\ 
     13 &         &          & 4.364565 &          & 0.444937 &  & 4.665748 &                  & 4.728925 \\ 
     14 &         &          & 0.964557 &          & 4.781066 &  & 0.000281 &                  & 3.157351 \\ 
     15 &         &          & 6.037635 &          & 5.429627 &  & 0.614126 &                  & 2.383873 \\ 
     16 &   &   & 5.975455 &   & 2.827826 &   & 6.176411 &                                     & 0.022179 \\ 
     17 &   &   & 0.159144 &   & 3.101400 &   & 3.698677 &                                     & 3.135448 \\ 
     18 &   &   & 6.194334 &   & 0.505015 &   & 1.349278 &                                     & 5.062797 \\ 
     19 &   &   & 1.518005 &   & 1.752444 &   & 5.651896 &                                     & 3.119667 \\ 
     20 &   &   & 2.570119 &   & 0.077924 &   & 0.048880 &                                     & 0.821506 \\ 
     21 &   &   & 2.836344 &   & 2.324862 &   & 1.604027 &                                     &  \\ 
     22 &   &   & 3.171423 &   & 5.525191 &   & 4.721183 &                                     &  \\ 
     23 &   &   & 4.005286 &   & 1.711153 &   & 0.000189 &                                     &  \\ 
     24 &   &   &   &   & 0.113893 &   & 1.563388 &                                            &  \\ 
     25 &   &   &   &   & 0.040828 &   & 3.165257 &                                            &  \\ 
     26 &   &   &   &   & 0.863436 &   & 0.025552 &   &   \\ 
     27 &   &   &   &   & 6.210510 &   & 3.165166 &   &   \\ 
     28 &   &   &   &   & 4.981590 &   & 1.566059 &   &   \\ 
     29 &   &   &   &   & 2.622501 &   & 0.897177 &   &   \\ 
     30 &   &   &   &   & 0.536382 &   & 5.028318 &   &   \\ 
     31 &   &   &   &   & 2.882208 &   & 6.282094 &   &   \\ 
     32 &   &   &   &   & 0.148144 &   & 2.400830 &   &   \\ 
     33 &   &   &   &   & 2.916385 &   & 3.127614 &   &   \\ 
     34 &   &   &   &   & 5.971181 &   & 2.312352 &   &   \\ 
     35 &   &   &   &   &   &   & 5.047211 &   &   \\  \cline{1-10} 
\end{tabular}
\caption{Angles (in radians) defining the $Z(\theta)$ gates in each of the circuits presented in the main text. \label{tab:angles}}
\end{table}

\section{Noise model process matrices}
\label{sec:proc_mat}
In this Appendix we list the process matrices and SPAM elements derived from GST experiments that define our error model for the 5-qubit device we demonstrate NACL on. These process matrices are completely-positive trace-preserving estimates of the corresponding operations.  (We note that, in order to estimate these process matrices, GST required that we also estimate the process matrix corresponding to the $Y(\pi/2)$ operation.  We omit that estimate here as our device model does not include the $Y(\pi/2)$ gate in the native gate set). All process matrices are given in the Pauli basis (\ie they are ``Pauli transfer matrices'') while the SPAM operations are given in the ``standard'' representation.  Because of throughput constraints only ``short'' GST circuits (\ie circuits for linear-inversion GST \cite{blume-kohout_demonstration_2017}) were used; each circuit was repeated 1024 times.

\begin{align}
	I &= \left(\begin{matrix}
		    1.0000  & -0.0000 &   0.0000 &   -0.0000 \\
    0.0042  &  0.9943  & -0.0064  &  0.0178 \\
   -0.0033  &  0.0120  &  0.9962  &  0.0186\\
    0.0029 &  -0.0182  & -0.0167  &  0.9928
	\end{matrix}\right) \nn\\
	X(\pi/2) &= \left(\begin{matrix}
	    1.0000  &  0.0000  &  0.0000  & -0.0000 \\
    0.0007  &  0.9988  & -0.0050  & -0.0055 \\
   -0.0010 &  -0.0060  &  0.0167  & -0.9980 \\
   -0.0017 &   0.0065  &  0.9979  &  0.0176
\end{matrix}\right) \nn \\
P_{0} &= \left(\begin{matrix}
     0.9997 &  -0.0006 \\
    0.0055   & 0.0231	
 \end{matrix}\right) \nn \\
P_{1} &= \left(\begin{matrix}
	    0.0003  &  0.0006 \\
   -0.0055 &   0.9769
\end{matrix}\right) \nn \\
\rho_0 &= \left(\begin{matrix}
    0.9903    &     0\\
         0   & 0.0097
          \end{matrix}\right)\nn.
\end{align}
Here, $P_{0}$ and $P_{1}$ are the imperfect POVM effects for projections onto the $\ket{0}$ and $\ket{1}$ states, respectively. $\rho_0$ is the density matrix for the single-qubit imperfect state preparation. Finally,

\begin{align}
CNOT = \left(\begin{smallmatrix}	
	1.000 &0 &0 &0 &0 &0 &0 &0 &0 &0 &0 &0 &0 &0 &0 &0 \\ 
0.012 &0.973 &0.016 &0.005 &0.005 &-0.002 &0.012 &-0.004 &-0.002 &0.003 &-0.004 &0.002 &-0.010 &0.008 &0.015 &-0.001 \\ 
0.001 &-0.009 &0.004 &-0.003 &-0.002 &0.000 &-0.023 &0.001 &-0.006 &-0.001 &-0.007 &0.003 &0.005 &-0.019 &0.974 &0.003 \\ 
0.002 &0.006 &0.000 &0.003 &-0.005 &-0.001 &0.002 &-0.021 &-0.010 &0.001 &0.003 &-0.010 &-0.001 &-0.007 &0.004 &0.983 \\ 
0.002 &0.001 &0.012 &-0.008 &0.015 &0.964 &0.017 &0.004 &0.001 &0.020 &-0.018 &0.003 &0.048 &0.020 &-0.002 &-0.004 \\ 
0.002 &-0.001 &0.004 &0.002 &0.980 &0.004 &-0.002 &-0.009 &0.018 &0.001 &-0.005 &0.012 &0.021 &0.042 &0.002 &0.005 \\ 
-0.002 &-0.003 &0.041 &0.002 &-0.009 &0.001 &0.005 &-0.018 &-0.005 &-0.002 &0.003 &0.977 &0.014 &-0.003 &0.000 &0.012 \\ 
-0.003 &-0.006 &-0.002 &0.045 &-0.006 &0.019 &0.015 &0.006 &-0.002 &0.022 &-0.968 &-0.001 &-0.006 &0.001 &-0.008 &0.005 \\ 
0.001 &0.007 &-0.004 &0.001 &0.000 &-0.019 &0.017 &-0.001 &0.011 &0.966 &0.019 &0.003 &0.012 &0.009 &-0.002 &-0.005 \\ 
0.001 &0.008 &0.004 &-0.001 &-0.021 &-0.000 &0.002 &-0.011 &0.981 &0.004 &-0.001 &-0.005 &0.014 &0.004 &0.002 &0.010 \\ 
-0.001 &-0.005 &0.007 &-0.002 &0.005 &0.005 &-0.003 &-0.975 &-0.011 &0.002 &0.007 &-0.020 &-0.003 &-0.002 &0.008 &-0.023 \\ 
-0.002 &-0.012 &0.004 &0.006 &0.003 &-0.021 &0.967 &0.001 &-0.005 &0.017 &0.016 &0.007 &0.003 &0.004 &0.021 &0.004 \\ 
-0.002 &-0.003 &-0.001 &0.001 &-0.021 &-0.035 &-0.008 &-0.001 &-0.010 &-0.006 &0.001 &-0.006 &0.987 &0.002 &0.001 &-0.000 \\ 
-0.008 &0.006 &0.012 &-0.001 &-0.043 &-0.020 &-0.003 &0.003 &-0.010 &-0.009 &0.003 &0.008 &0.011 &0.970 &0.016 &0.007 \\ 
0.005 &-0.018 &0.973 &0.003 &-0.004 &-0.009 &0.002 &0.008 &0.002 &0.005 &-0.001 &-0.039 &-0.004 &-0.007 &0.005 &-0.005 \\ 
0.000 &-0.007 &0.005 &0.982 &0.005 &0.002 &-0.008 &0.003 &0.003 &-0.009 &0.040 &0.002 &0.002 &0.005 &0.001 &0.001 \\
    \end{smallmatrix}\right)\nn
\end{align}

We also list below various error metrics for these noisy operators (as compared to ideal operators).

\begin{table}[hpb!]
    \centering
    \begin{tabular}{|c||c|c|}
    \hline
        Gate label & Infidelity & 1/2 diamond distance\\
    \hline
        $I$ & $2.8\cdot10^{-3}$& $1.7\cdot10^{-2}$ \\
    \hline
        $X(\pi/2)$ & $8.8\cdot10^{-4}$ &$1.1\cdot10^{-2}$\\
    \hline
        $CNOT$ & $1.9\cdot10^{-2}$& $5.0\cdot10^{-2}$\\
    \hline
    \hline
        $\rho_0$ & $9.7\cdot10^{-3}$ & - \\
    \hline
        $P_0$ & $2.0\cdot10^{-3}$ & - \\
    \hline
        $P_1$ & $2.3\cdot10^{-2}$ & - \\
    \hline
    \end{tabular}
    \caption{Error metrics for noisy operations (compared to ideal operations) used in our device model input to NACL.  For gate operations, entanglement infidelity and diamond distance are presented, while for SPAM operations, only state infidelity is used.}
    \label{tab:error_metrics}
\end{table}

``Infidelity'' for gate operations is taken to be average gate infidelity, \emph{i.e.,} $1-\bar{F}$, where $\bar{F}$ is the average gate fidelity (with respect to the desired target operation), as defined in Eq. \eqref{eq:Fbar_process}.  For SPAM operations we simply use state infidelity, i.e., 
\begin{equation}
1-F(\rho , \sigma) = 1-\Big(\Tr \sqrt{\sqrt{\rho} \sigma \sqrt{\rho} }\Big)^2
\end{equation}
Half-diamond distance, denoted $\epsilon_\diamond$, is defined as \begin{equation}
    \epsilon_\diamond(A,B) = \tfrac{1}{2}||A-B||_\diamond=\tfrac{1}{2}\sup_\rho||\left(A\otimes\mathbb{1}_d[\rho]\right) - \left(B\otimes\mathbb{1}_d[\rho]\right)||_1,
\end{equation}
where $||\cdot||_1$ is the trace norm, $\sup$ is taken over all density matrices of dimension $d^2$, and $d=\dim A=\dim B$.  

Average gate infidelity may be thought of as, averaged over the Haar measure, the infidelity of a state that has passed through the gate's channel; diamond distance may be thought of as a worst-case error rate.  Average gate infidelity is quadratically more sensitive to stochastic error than unitary error, while diamond distance is equally sensitive to both classes of errors~\cite{sanders2015bounding}.

\section{Noise model for a trapped-ion quantum computer}
\label{sec:eff_nm}

In addition to the noise model presented in the main text, we also ran NACL using an additional noise model, that is an effective model formed from error metrics derived from a near-term trapped-ion quantum computer. We adapt the coarse-grained error maps used to model errors during execution of a common trapped-ion gate set developed by Trout \etal in Ref. \cite{trout_simulating_2018}. In particular, the native gates in the processor are assumed to be $X(\theta), Y(\theta), Z(\theta)$ and $XX(\theta) \equiv e^{i\theta X\otimes X}$, where the first three are single qubit rotations about the three orthogonal axes and the last is an arbitrary angle Molmer-S{\o}rens{\o}n interaction between two qubits. The quantum channels representing the noisy versions of each of these gates are given by:
\begin{align}
    \mathcal{E}_X(\theta) &= \mathcal{D}(p_{\rm d}) \circ \mathcal{W}(p_{\rm dep}) \circ \mathcal{R}_X(p_\alpha) \circ \mathcal{U}_X(\theta), \nn \\
    \mathcal{E}_Y(\theta) &= \mathcal{D}(p_{\rm d}) \circ \mathcal{W}(p_{\rm dep}) \circ \mathcal{R}_Y(p_\alpha) \circ \mathcal{U}_Y(\theta), \nn \\
    \mathcal{E}_Z(\theta) &= \mathcal{D}(p_{\rm d}) \circ \mathcal{W}(p_{\rm dep}) \circ \mathcal{R}_Z(p_\alpha) \circ \mathcal{U}_Z(\theta), \nn \\
    \mathcal{E}_{XX}(\theta) &= [\mathcal{D}_1(p_{\rm d,1})\otimes \mathcal{D}_2(p_{\rm d,2})] \circ \nn \\ & ~\quad [\mathcal{W}_1(p_{\rm dep})\otimes  \mathcal{W}_1(p_{\rm dep})]  \circ \nn \\ & ~\quad
    \mathcal{H}(p_{\rm xx}) \circ \mathcal{H}(p_{\rm h}) \circ \mathcal{U}_{XX}(\theta). \nn
\end{align}
Here, $\mathcal{U}_k(\theta)$ represents an ideal rotation about axis $k$ (\eg $\mathcal{U}_X(\theta)\rho = e^{-i\theta X}\rho e^{i\theta X}$), $\mathcal{R}_k(p_\alpha)$ represents the effects of rotation angle imprecision about axis $k$ (\eg $\mathcal{R}_X(p_\alpha)\rho = (1-p_\alpha)\rho + p_{\alpha}X\rho X$),  $\mathcal{W}(p_{\rm dep})$ is a depolarizing channel (\ie $\mathcal{W}(p_{\rm dep})\rho = (1-p_{\rm dep})\rho + p_{\rm dep}I$), $\mathcal{D}(p_{\rm d})$ is a dephasing channel (\ie $\mathcal{D}(p_{\rm d})\rho = (1-p_{\rm d})\rho + Z\rho Z$, and finally, $\mathcal{H}(p)\rho = (1-p)\rho + XX\rho XX$, is a two-qubit channel that represents the effects of an imprecise rotation (when $p=p_{\rm xx}$) or the effects of ion heating (when $p=p_{\rm h}$). The subscripts on any of these channels (in the case of the two-qubit operation) denotes action on that qubit.

In addition to these imperfect gates, we model SPAM errors by following an ideal ground state preparation with a depolarizing channel, and by preceding ideal single qubit measurement POVM effects by a depolarizing channel, \ie
\begin{align}
    \langle\langle 0 \vert &\rightarrow \langle\langle 0 | \mathcal{D}(p_{\rm dep}) \nn \\
    \vert i \rangle\rangle  &\rightarrow \mathcal{D}(p_{\rm dep}) \vert i \rangle\rangle \quad\quad \textrm{for} ~~ i=0,1,
\end{align}
where we have notated state preparation and measurement effects as Hilbert-Schmidt vectors. Finally, in order to capture noise during idle cycles, all idles are modeled as a depolarizing channel $\mathcal{D}(p_{\rm idle})$.

This effective noise model captures many of the non-idealities in typical ion trap quantum computing architectures. However, note that under this model there are no connectivity restrictions and it is possible to perform a two-qubit gate between any two of qubits. In the following computations we use the error rates:
\begin{align}
    p_{\rm d} &= 1.5\times 10^{-4} \nn \\
    p_{\rm dep} &= 8\times 10^{-4} \nn \\
    p_{\rm d,1} = p_{\rm d,2} &= 7.5\times 10^{-4} \nn \\
    p_\alpha &= 1\times 10^{-4} \nn \\
    p_{\rm xx} &= 1\times 10^{-3} \nn \\
    p_{\rm h} & = 1.25\times 10^{-3} \nn \\
    p_{\rm idle} & = 8\times 10^{-4}
\end{align}

\section{NACL training times} \label{sec:training_time}

In this Appendix we discuss computational resources that are needed to compile circuits with NACL. NACL is initialized with random quantum circuit. That circuit is evolved in time such that it minimizes properly defined cost function. This aspect of our approach together with non-convex optimization landscape with many local minima make it hard to define ``compilation time'' or ``time to solution''. Furthermore, the optimization could be terminated before reaching global minimum of the cost function. As we show here, such imperfect compilations are still better than other compilations in many cases.

\begin{figure}
\includegraphics[width=\textwidth]{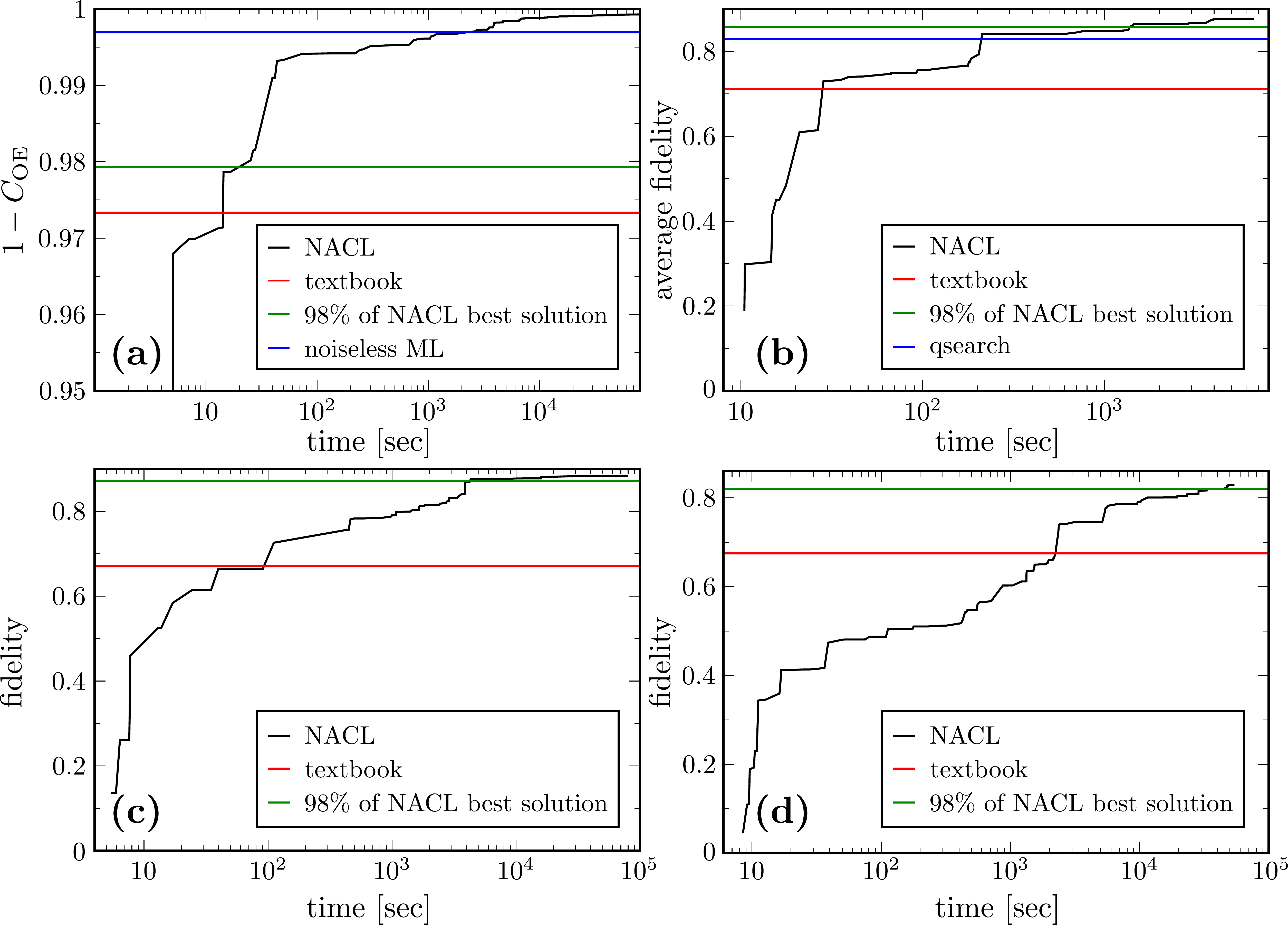}
\caption{NACL performance. Panels show compilation quality as a function of wall-clock time. The definition of compilation quality depends on a particular task, see text for details. Panels (a), (b), (c) and (d) show NACL performance as applied to observable extraction, 3-qubit QFT compilation, 4-qubit and 5-qubit $W$ state preparation, respectively.
}
\label{fig:comp_time}
\end{figure}

Instead of working with a particular definition of ``compilation time'', we will analyze how quickly NACL can reduce the cost and find circuits that perform better than textbook ones. This gives more information about performance of the algorithm than single number that characterizes compilation time. Our results are summarized in Fig.~\ref{fig:comp_time}. It shows all use cases that we consider in the main text. Panel (a) shows how quality of the compilation is increased in time. Here (as well as in other panels) time indicates wall-clock time, measured in seconds. Compilation quality is measured as $1 - C_{\textrm{OE}}$, where $C_{\textrm{OE}}$ is the observable extraction cost function defined in Eq.~\eqref{eq:obs_ext_cost}. We use the same training data set as the one considered in the main text. The results suggest that NACL quickly finds an algorithm that outperforms the textbook one. It took NACL only 15 seconds to find a circuit that has a lower cost function than the textbook one. Further improvements however are more costly and reaching global minimum took more than 20 hours. As expected, one is dealing with diminishing returns in running longer optimizations, as the algorithm that reached 98\% of the best achievable quality was found only after 26 seconds. 

The above cost analysis is representative for other applications. Panel (b) shows results for 3-qubit QFT considered in Sec.~\ref{sec:impl_comp}. The plot displays how average fidelity of the channel $\mathcal{E}$ defined by the noisy circuit is improved as a function of time. Here, the fidelity is defined by $(dF_e(\mathcal{U}^\dagger \circ \EC)+1)/(d+1)$, where $d = 8$ for a three qubit example, $F_e$ is the entanglement fidelity defined in Sec.~\ref{sec:CostFunctions} and $\mathcal{U}$ is the ideal channel implementing 3-qubit QFT. Again, we find that NACL quickly surpasses textbook solution but takes significantly more time to find the global optimum. Panels (c) and (d) present results for $W$ state preparation on 4 and 5 qubits respectively. Here, the compilation quality is defined as a fidelity between the noisy state and exact $W$-state. As expected, compiling bigger circuits is more expensive due to exponential scaling of quantum simulation. Nevertheless, NACL manages to surpass best textbook algorithm within 110 seconds and 37 minutes for 4- and 5-qubit $W$-state preparation, respectively.

Table~\ref{tab:comp_times} combines compilation times (as defined by time to beat textbook algorithm and reaching 98\% of best NACL result) for various applications of NACL. We used 16-core desktop computer to generate the data shown in Table~\ref{tab:comp_times} and Fig.~\ref{fig:comp_time}.

\begin{table}
\centering
\begin{tabular}{|c||c|c|}
\hline
algorithm & time to beat textbook alg. & time to achieve 98\% of best NACL \\
\hline
observable extraction & 15 sec & 26 sec \\
\hline
3-qubit QFT & 28 sec & 23 min \\
\hline
4-qubit QFT & 110 sec & 1.3 h \\
\hline
5-qubit QFT & 37 min & 12 h \\
\hline
\end{tabular}
\caption{Compilation times for all NACL applications considered in the main text. The data presented in the table is extracted from Fig.~\ref{fig:comp_time} and shows time needed to surpass textbook algorithm and time to reach 98\% of best circuit found by NACL.
}
\label{tab:comp_times}
\end{table}

Compilation times presented in Table~\ref{tab:comp_times} and Fig.~\ref{fig:comp_time} may seem to be long. We would like to stress however that in the NISQ era, compilation quality is more important that time in which that compilation was found. One is trying to use  quantum computers to its fullest capability and avoid noise as much as possible. We envision that NACL will be used to create libraries of commonly used algorithmic routines for specific devices. It is worth investing time on classical computers such that every quantum algorithm built using precompiled routines performs better than the one obtained with just-in-time compilers.

\end{document}